# Topological state transitions in electromagnetic topological defects

Peng Shi[1,*], Qiang Zhang[1], Xiaocong Yuan[2,1*]

[1]*Nanophotonics Research Centre, Institute of Microscale Optoelectronics & State Key Laboratory of Radio Frequency Heterogeneous Integration, Shenzhen University, Shenzhen 518060, China*
[2]*Zhejiang Lab, Research Center for Humanoid Sensing, Research Institute of Intelligent Sensing, Hangzhou 311100, China*

[*]Authors to whom correspondence should be addressed: *pittshiustc@gmail.com*, and *xcyuan@szu.edu.cn*

**Abstract:** The recent emergence of electromagnetic topological defects has attracted wide interest in fields from topological photonics to deep-subwavelength light-mater interactions. Previously, much of the research has focused on constructing specific topological defects but the fundamental theory describing the physical mechanisms underlying their formation and transitions is lacking. Here, we present a spin–orbit coupling based theory describing such mechanisms for various configurations of spin topological defects in confined electromagnetic fields. The results reveal that their formation originates from the conservation of total angular momentum and that their transitions are determined by anisotropic spin–orbit couplings. By engineering the spin–orbit couplings, we observe the formation and transitions of Neél-type, twisted-type, and Bloch-type spin topological defects in confined electromagnetic fields. A stable Block-type spin topological defect is reported for the first time. Our theory can also describe the transitions of field topological defects. The findings enrich the portfolio of electromagnetic topological defects, deepen our understanding of conserved laws, spin–orbit couplings and transitions of topological defects in confined electromagnetic systems, and predict applications in high-density optical data transmissions and chiral quantum optics.

## 1. Introduction

Topological defects, which are domains of discontinuity in order-parameter fields, play an important role in modern physics and biology [1,2]. Therein, skyrmions, which are topological defects in chiral magnets [3], are considered a promising route toward high-density information storage and transfer [4–6]. Typically, two configurations of skyrmions are observed [6]: Néel-type skyrmions, in which spins rotate in cycloidal spirals, and Bloch-type skyrmions, in which spins rotate in helical spirals. If the spins of the skyrmion state simultaneously rotate cycloidally and helically, a twisted skyrmion is formed [7]. Recently, topological defects have been predicted in electromagnetic (EM) systems and various EM topological defects have been demonstrated [8–17], particularly in confined fields, including Néel-type fields and spin skyrmions [18–23], spin quasiparticles [24], twisted-type spin skyrmions [25], Néel-type fields and spin merons lattices [26–28], and Néel-type spin domain walls [29], all of which have potential applications in optical manipulation [30], chiral sorting [31–33], data storage [34], ultrafast deep-subwavelength imaging [35], and chiral quantum optical technologies [36–38].

In condensed matter physics, the magnetic skyrmion is a topologically nontrivial spin that forms via spin–orbit coupling (SOC) [6]. The transitions among Néel-type, Bloch-type, and twisted-type configurations [39,40], as well as the transformations between skyrmion, antiskyrmion, and meron topologies [41,42] are readily described using SOCs. In EM field, much of the previous research has primarily focused on the formation of topological defects from separate aspects, whereas the influence of SOCs on the transitions of topological states has been ignored [43,44]. This confuses readers regarding the mechanisms underlying the formation of various configurations of EM topological defects, which will face challenges in constructing some configurations of EM topological defects. For example, no stable Bloch-type EM topological defect has been formed to date, although researchers have employed chirality to twist spin vectors and construct Bloch-type topologies in the single center plane of chiral material [45,46].

In this work, we propose a SOC-based theory to describe the mechanisms behind the formation and transitions of various configurations of EM spin topological defects in confined fields. By considering the conserved properties of an EM system, we construct a variational equation to explain the formation of various configurations of spin topological defects; moreover, the transitions of these configurations are governed by the intrinsic SOC derived from the optical Dirac equation, in which the real part is given by the gradient of the EM helical density and the imaginary part represents the EM transverse spin. Then, by engineering the SOCs in a multilayered structure, we construct the twisted-type, Néel-type, and Bloch-type spin topological defects. A stable Bloch-type spin defect was constructed with a confined EM field for the first time. Furthermore, we show that if the dispersion relation of the EM modes crosses over the light cone, a photonic analog of the topological phase transition occurs. In addition, the theory also describes the formation and transitions of various configurations of electric field topological defects in EM system. Our theory is general and describes the transitions of spin and field topological defects universally, and the present theoretical framework is important in the understanding of topological defects for all classical wave fields.

## 2. Results

We consider EM spin topological defects in an air–medium–air structure excited by monochromatic circularly polarized light [see Fig. 1(a), inset]. The results generalize to arbitrary multilayered structures because the expressions in each layer are similar. Such multilayered structures are beneficial in realizing refractive-index-engineered artificial metamaterials [47,48] and have also been widely studied in fields concerning acoustic waves [49], sound waves [50], and elastic waves [51]. The spin–orbit properties of topological defects are described by the optical Dirac equation [52]:

$$\hat{\mathbf{H}}|\Psi\rangle = v\hat{\boldsymbol{\alpha}} \cdot \hat{\mathbf{p}}|\Psi\rangle = i\hbar \frac{\partial}{\partial t}|\Psi\rangle, \quad (1)$$

where $|\Psi\rangle$ denotes the Riemann–Silberstein vector [53], $v = 1/\sqrt{\varepsilon\mu}$ the velocity of light in a medium, permittivity $\varepsilon$ and permeability $\mu$, $\hat{\mathbf{p}}$ the momentum operator, $\hat{\mathbf{H}}$ the Hamiltonian operator [54], and $v\hat{\boldsymbol{\alpha}} = v[\mathbf{0}, \hat{\mathbf{S}}; \hat{\mathbf{S}}, \mathbf{0}]$ (with $\hat{\mathbf{S}}$ the spin-1 matrix in SO(3) [55]) the operator describing the transportation of photons. From this equation, one determines that the conserved properties of the spin angular momentum (SAM) and orbital angular momentum (OAM) are given by (see Supplementary Material II for details):

$$\dot{\hat{\boldsymbol{\Sigma}}} = \frac{i}{\hbar}\left[\hat{\mathbf{H}}, \hat{\boldsymbol{\Sigma}}\right] = -v\hat{\boldsymbol{\alpha}} \times \hat{\mathbf{p}}, \quad (2)$$

and

$$\dot{\hat{\mathbf{L}}} = \frac{i}{\hbar}\left[\hat{\mathbf{H}}, \hat{\mathbf{L}}\right] = v\hat{\boldsymbol{\alpha}} \times \hat{\mathbf{p}}, \tag{3}$$

where the over-dot signifies time differentiation. Here, the SAM operator $\hat{\boldsymbol{\Sigma}} = \hbar[\hat{\mathbf{S}}, \mathbf{0}; \mathbf{0}, \hat{\mathbf{S}}]$ and the OAM operator $\hat{\mathbf{L}} = \mathbf{r} \times \hat{\mathbf{p}}$ are not conserved separately in the system. However, for a multilayered system, the total angular momentum (AM) is conserved in the normal direction. This conserved property originates from the symmetry of the EM system [28], and thus the conservation of the total AM is protected from variations in the local spin vectors in the formation of EM spin topological defects. Therefore, the variational equation,

$$\delta \langle \mathbf{J} \cdot \mathbf{J} \rangle = 0, \tag{4}$$

can explain the formation of various configurations of EM spin topological defects by setting the local spin vector as [6]: $\boldsymbol{\Sigma} = \hbar\sigma_h[\hat{\mathbf{r}}\sin\Theta\cos\Phi + \hat{\boldsymbol{\varphi}}\sin\Theta\sin\Phi + \hat{\mathbf{z}}\cos\Theta]$, where $\sigma_h$ denotes the EM helicity of the incident field, and $\Theta$ (the polar angle) and $\Phi$ (the azimuthal angle) describe the vector direction of the local spin vector. Note that the Dirac bracket notation in Eq. (4) represents an integral in the horizontal plane. Generally, both angles depend on the horizontal coordinates of the multilayered system. Here, for simplicity, we set $\Phi = \text{constant}$ and $\Theta = \Theta(r)$ with $r$ the polar coordinate. By solving the variational equation (4), one finds the nontrivial solution gives (see Supplementary Material III for details): 1. when $\Phi = 0$, a Néel-type skyrmion, 2. when $\Phi = \pi/2$, a Bloch-type skyrmion, and 3. when $\Phi = \pi/4$, a twisted-type skyrmion. These results show that various configurations of EM spin topological defects form in a multilayered structure and the formation of these defects originates from the conserved properties of the EM system.

Next, to understand the mechanism behind transitions between various configurations of EM spin topological states, we consider the SOCs as in the condensed matter physics approach. For magnetic skyrmion systems, a Bloch-type magnetic skyrmion stabilizes in chiral magnets through the Dresselhaus SOC, whereas a Néel-type magnetic skyrmion arises in polar magnets through the Rashba SOC [40]. In the EM system, from the conserved properties of the optical Dirac equation, the SOC (or spin–orbit interaction) is described by operator $v\hat{\boldsymbol{\alpha}} \times \hat{\mathbf{p}}$. By calculating the inner product of this operator, one obtains the real and imaginary parts of this SOC term, denoted $\mathbf{H}_{\text{SO}}$, given as (see Supplementary Material II for details)

$$\text{Re}\, \mathbf{H}_{\text{SO}} = -\frac{\hbar\omega}{k^2}\nabla C, \tag{5}$$

and

$$\text{Im}\, \mathbf{H}_{\text{SO}} = \frac{\hbar\omega^2}{2k^2}\nabla \times \mathbf{P}^A. \tag{6}$$

One finds that the SOCs are therefore related to the helical density $C$ and the kinetic Abraham–Poynting momentum $\mathbf{P}^A$ [56]. Therein, variations in EM helicity correspond to variations in longitudinal spin ($\mathbf{S}_l = \sum \hbar\sigma_h\hat{\mathbf{k}}$ by expressing the structured light as a superposition of basis plane waves [57]). Thus, this SOC effect is found ubiquitously in spin–orbit interaction phenomena such as the spin Hall effect of light and the spin-to-orbital AM conversion of light [43,44]. Meanwhile, the vorticities associated with $\mathbf{P}^A$ correspond to EM transverse spins [28,58], which are related to the structural properties of EM fields [15,59]. Moreover, the EM transverse spin is also involved in the intrinsic spin-momentum locking of light [28,57,58,60] and found widely in spin–orbit interaction phenomena such as unidirectional guided waves of light [36–38], orbital-to-spin AM conversions [43], and photonic topological defects [19,21–28].

To explain the inherent relationship between the SOC term and the configurations of EM spin topological defects in more detail, we consider the example of photonic spin topological defects displaying $C_4$ rotational symmetry [26–28] within a medium layer, in which the permittivity is described by a Lorentz model (see Supplementary Material I for details), $\varepsilon(\omega) = \varepsilon_0[1 - \omega_p^2/(\omega^2 - \omega_i^2 + i\omega\gamma)]$. Here, $\omega_p$ denotes the plasma frequency, $\omega_i$ the displacement electron resonance frequency, and $\gamma = 1/\tau$ the characteristic frequency with $\tau$ the relaxation time of the electron gas.

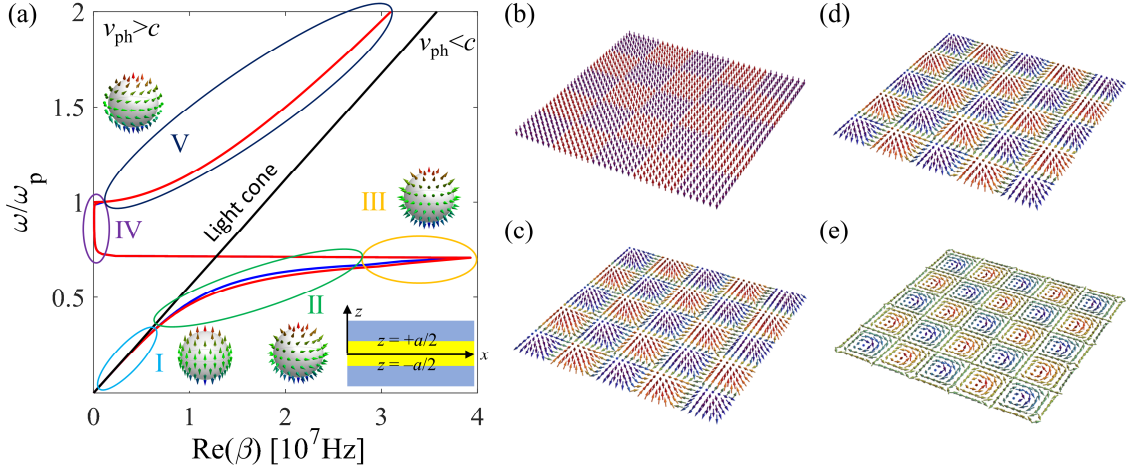

**Fig. 1.** EM topological defects in the air–medium–air structure: (a) Dispersion relations of the EM modes in an air–medium–air structure (bottom right inset); see Supplementary Material IV for details. (b) At low frequencies (region I, $\omega \ll \omega_p$), the spin topological state contains purely spin 'up' and spin 'down' states. (c) As the frequency increases (region II), the complex constant of propagation for the lossy mode leads to a small kinetic Abraham–Poynting momentum in the $z$-direction, and thus a nonzero azimuthal SAM, which manifests as a twisted-type spin topological defect; red/blue lines indicate symmetric/anti-symmetric surface modes. (d) In the lossless limit (region III), a Néel-type spin topological defect is present. As the frequency increases further (region IV), a band gap appears because the constant of propagation is purely imaginary. (e) At frequencies above $\omega_p$ (region V), a Bloch-type spin topological defect forms through the cancellation of the radial SAM from the intrinsic spin-momentum locking property of light. Here, the skyrmion number for the topological spin defects is ±1/2; see Supplementary Material VI for details.

In the low-frequency region $\omega \ll \omega_p$ [Fig. 1(a), region I], the relative permittivity gives $\text{Re}\{\varepsilon(\omega)\} = 1 - \omega_p^2/(\omega^2 + \gamma^2) \ll 0$ and $\text{Im}\{\varepsilon(\omega)\} = \omega_p^2/(\omega^3/\gamma + \omega\gamma) \gg 0$. The material is metallic and only the transverse magnetic (TM) surface plasmon polariton (SPP) modes are excited in the structure ($H_z = 0$). The constant of propagation for the SPP mode is $\beta = k_0\sqrt{\varepsilon(\omega)/(\varepsilon_0 + \varepsilon(\omega))} \approx k_0$ with $k_0$ the wavenumber in air. In the medium layer, the wavevector in the $z$-direction, which also represents the decaying factor, is $k_z^m = (\sqrt{\beta^2 - (k^m)^2} \gg \beta$ and thus the EM field is localized at the structure's surface. In this instance, the $z$-component of the real part of the SOC term (Re $\mathbf{H}_{SO})_z$, which describes the variation in helical density of the $z$-component, is much larger than the horizontal component (Re $\mathbf{H}_{SO})_r$. Meanwhile, the imaginary part of the SOC term (Im $\mathbf{H}_{SO}$) affects the horizontal SAMs via two mechanisms [40]: 1. Rashba-like SOC leading to a radial SAM through $(\hat{\mathbf{n}} \times \mathbf{P}_\varphi^A) \cdot \mathbf{S}$, which is also associated with the intrinsic spin-momentum locking of EM field [60], and 2. Azimuthal SAM originating from the $z$-component of $\mathbf{P}^A$ through $(\nabla_r \times \mathbf{P}_z^A)$ [58]. The configuration of the EM spin topological defects is determined by these two terms simultaneously. Here, $\hat{\mathbf{n}}$ denotes the outward normal direction and $\mathbf{S}$ the dispersive SAM [56]. In this instance ($\omega \ll \omega_p$), the effects from both mechanisms are extremely small and therefore we assert the SOC occurs only in the *z*-direction, implying that a *z*-component longitudinal spin is prominent (see Supplementary Material V for details). In the low-frequency limit, the spin topological defects may be considered universally to be a combination of purely 'up' and 'down' states [Fig. 1(b)].

As the frequency increases into the near-infrared or visible band [Fig. 1(a), region II], though the material maintains its metallic properties, the symmetric and antisymmetric EM modes are excited within the structure, both of which are lossy and have constants of propagation containing real and imaginary parts. The *z*-component wavenumber $k_z^m$ is of the same order of magnitude as the constant of propagation. If we consider either one of these modes alone, both the real and imaginary parts of the *z*-component SOC term are nonzero and thus there is a *z*-component of SAM (From [28], in the layer, the *z*-component SAM contains the longitudinal spin related to the EM helicity and transverse spin simultaneously.). The azimuthal kinetic Abraham–Poynting momentum, by contrast, results in a radial SOC as well as the radial SAM through the term $(\hat{\mathbf{n}} \times \mathbf{P}_\varphi^A) \cdot \mathbf{S}$. Because of the complex constant of propagation for each of these EM modes, the kinetic Abraham–Poynting momentum in the *z*-direction is also nonzero, and thus the azimuthal SAM component appears for $S_\varphi \propto (\nabla_r \times \mathbf{P}_z^A)$ (see Supplementary Material V for details). The spin topological defects are categorized as twisted-type meron lattices [Fig. 1(c)]. Note that the azimuthal SAM is smaller than the radial SAM by an order of magnitude but this azimuthal SAM and the resulting twists can be tuned using the lossy property (the characteristic frequency *γ*) of the EM mode [61].

As the frequency increases further when the loss can be ignored [Fig. 1(a), region III], the helical density vanishes in the layer and the real part of $\mathbf{H}_\text{SO}$ is also zero. However, the *z*-component of the imaginary part of the SOC term (Im $\mathbf{H}_\text{SO})_z$ is nonzero because of the inhomogeneities of the EM mode in the horizontal plane. We must emphasize that the imaginary part of the SOC term gives rise to transverse spin but the real part of the SOC term only represents the variation in longitudinal spin, which exists and is homogeneous when Re $\mathbf{H}_\text{SO}$ = 0 [28]. Thus, there is a *z*-component of SAM. Otherwise, there is only an azimuthal kinetic Abraham–Poynting momentum in the layer and thus a radial SAM related to $(\hat{\mathbf{n}} \times \mathbf{P}_\varphi^A) \cdot \mathbf{S}$ appears and the azimuthal SAM given by $(\nabla_r \times \mathbf{P}_z^A)$ vanishes (see Supplementary Material V for details). In this instance, by ignoring the loss in this lossless limit, the configuration of the spin topological defect is transformed from twisted-type to Néel-type [Fig. 1(d)].

As the frequency increases further with $\omega < \omega_p$ [Fig. 1(a), region IV], Re$\{\varepsilon(\omega)\} < 0$ and Re$\{\varepsilon_0 + \varepsilon(\omega)\} > 0$, which leads to a purely imaginary constant of propagation. In this instance, propagating modes are forbidden and a photonic band gap appears (see Supplementary Material IV for details).

Subsequently, when the frequency is larger than the plasma frequency, $\omega > \omega_p$ [Fig. 1(a), region V], the real part of the relative permittivity obeys $0 < \text{Re}\{\varepsilon(\omega)/\varepsilon_0\} \approx 1 - \omega_p^2/\omega^2 < 1$. The medium changes from being metallic to dielectric and a symmetric slot-waveguide-like mode (radiation mode) is present ($E_z/H_z = i\sqrt{\mu_0/\varepsilon_0}$) [62]. In this instance, the constant of propagation for the symmetric TM and transverse electric (TE) modes are identical (see Supplementary Material IV for details). Because the normal electric field component is dominant in the layer, the effective index for this slot-waveguide-like mode is $n_\text{eff} \approx \sqrt{\varepsilon(\omega)/\varepsilon_0}$. This ensures that the wavevector in the *z*-direction, given by $k_z = \sqrt{\beta^2 - k^2}$ with constant of propagation $\beta = k_0 n_\text{eff}$ and wavenumber of the medium $k = k_0\sqrt{\varepsilon(\omega)/\varepsilon_0}$, is extremely small. Thus, the helical density is homogeneous along the *z*-direction in the layer and (Re $\mathbf{H}_\text{SO})_z$ = 0. Thus, the longitudinal spin is homogeneous through the layer and the variation in the z-component of SAM is determined by the imaginary part of the SOC term (Im $\mathbf{H}_{SO}$)$_z$. In addition, the kinetic Abraham–Poynting momentum in the z-component is nonzero, and the azimuthal SAM appears as $S_\varphi \propto (\nabla_r \times \mathbf{P}_z^A)$. Moreover, one can understand the behavior of the radial SAM through the intrinsic spin-momentum locking property of light by supposing that this SAM component is produced by a superposition of individual SAMs in the upper and lower interfaces. Because the wavevector in the z-direction $k_z$ is extremely small in the layer, the individual radial SAM in the upper and lower interfaces can be regarded as homogeneous through the medium layer. Moreover, because the outward normal direction of the upper interface is inverse to that of the lower interface, the direction of the vectors for the individual radial SAMs in the upper and lower interfaces, which are given by the radial SOC $(\hat{\mathbf{n}} \times \mathbf{P}_\varphi^A) \cdot \mathbf{S}$, are opposite. In this way, the radial SAM is canceled out in the layer (see Supplementary Material V for details). This is a manifestation of a Bloch-type spin topological defect [Fig. 1(e)]. Note here that these defects are stable, and their topology is maintained throughout the layer.

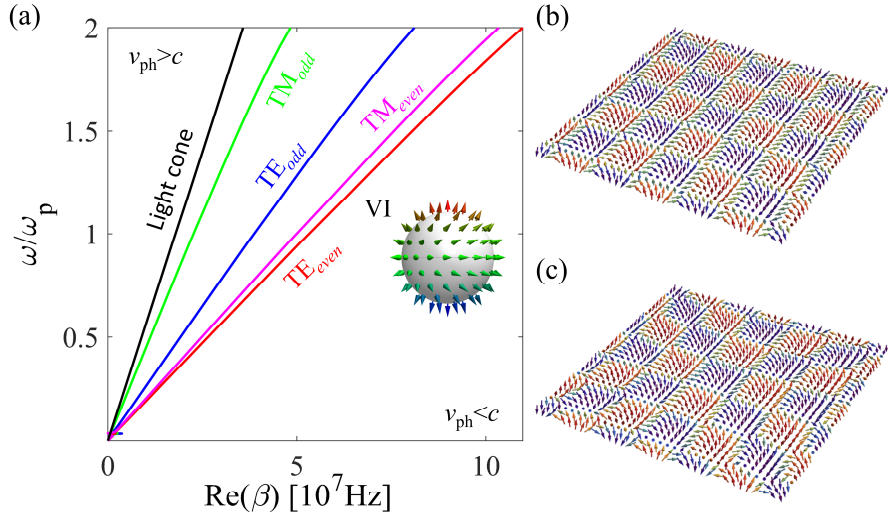

**Fig. 2.** EM topological defects in the air–dielectric–air structure: (a) Dispersion relations of the EM modes in the air–dielectric–air structure (see Supplementary Material IV for details). (b) The twisted-type spin defect is formed with either symmetric or antisymmetric modes separately. The skyrmion number of this spin topological defect is ±1/2 (see Supplementary Material VI for details). (c) When both symmetric and antisymmetric modes are excited simultaneously within the structure, no connected boundaries of the spin defects are seen and therefore the topologies of the spin textures are trivial, indicating that a photonic analog of the topological phase transition occurs. This transition also occurs between the region beyond the light cone (region V) as well within the light cone [Fig. 1(a), regions II and III].

In addition, for ordinary dielectrics for which the imaginary part of the permittivity is negligible and $\omega_i$ is taken into account, the relative permittivity is Re{$\varepsilon(\omega)/\varepsilon_0$} > 1 [Fig. 2(a), region VI], and four waveguide modes are excited in the structure [Fig. 2], namely, the symmetric TE (TE$_{even}$), symmetric TM (TM$_{even}$), antisymmetric TE (TE$_{odd}$), and antisymmetric TM (TM$_{odd}$) modes (see Supplementary Material IV for details). If we consider solely the symmetric or antisymmetric modes, the real part of the radial and normal SOC terms and the imaginary part of the radial, azimuthal, and normal SOC terms exist simultaneously. Thus, the radial, azimuthal, and normal SAM components are present simultaneously and manifest as a twisted-type spin topological defect [Fig. 2(b)].

From Fig. 2(a), the symmetric and antisymmetric modes occur with differing constants of propagation within the light cone. If one considers the spin topological defects constructed by the hybrid modes containing both symmetric and antisymmetric modes, the topologies are trivial because one cannot find a connected boundary to define the topological number [Fig. 2(c)]. In contrast, for the slot-waveguide-like mode, only the symmetric modes exist and moreover, the constant of propagation for these symmetric TE and TM modes are identical. Thus, the topology of the spin topological defect constructed from the slot-waveguide-like modes is maintained. Moreover, because the SPP mode or waveguide mode is localized inside the light cone ($v_{\text{ph}} < c$) whereas the slot-waveguide-like mode is outside ($v_{\text{ph}} > c$), a photonic analog of topological phase transition occurs as the EM modes cross over the light cone [63,64], i.e., the phase velocity changes from subluminal to superluminal.

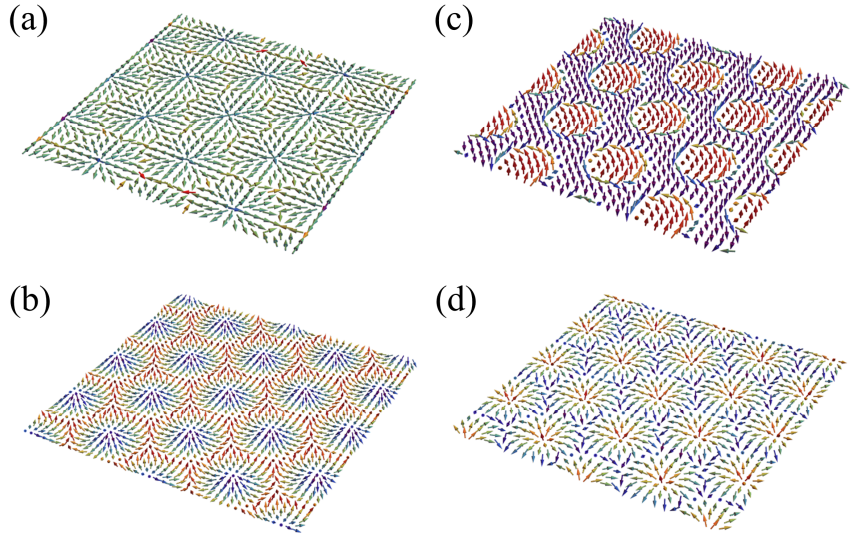

**Fig. 3.** Field topological defects in the air–medium–air structure. In region I [Fig. 1(a)], the horizontal electric field components are much larger than the normal electric field component in the layer and no field topological defect (a) exists for the skyrmion number $N_{SK} = 0$. In regions II and III [Fig. 1(a)], the azimuthal electric field component is absent and a Néel-type skyrmion (b) forms. In region V [Fig. 1(a)], the radial electric field component disappears through mode coupling, and a Bloch-type skyrmion (c) forms. Finally, in region VI [Fig. 2(a)], a twisted skyrmion (d) forms. The skyrmion number of the unit cell of skyrmion lattice is ±1 (see Supplementary Material VI for details).

In addition, our theoretical framework also accounts for the transitions of topological defects constructed by electric field distributions. Here, we take the electric field skyrmion defects displaying $C_6$ rotational symmetry as an example [18,20]. At low frequencies [Fig. 1(a), region I], the horizontal electric field components are much larger than the normal component for $k_z^m \gg \beta$ (from $\nabla \cdot \mathbf{E} = 0$) [Fig. 3(a)]. In the low frequency limit, there are primarily horizontal electric field components and the topology of this field texture is trivial because the skyrmion number of this field texture is considered to be zero. Then, in regions II and III [Fig. 1(a)], the material possesses metallic properties and only the TM polarized SPP modes are excited in the structure. In this instance, the azimuthal electric field components are absent, which is a manifestation of the Néel-type topological defect [Fig. 3(b)]. Subsequently, in region V [Fig. 1(a)], the slot-waveguide-like modes are excited, and the radial electric field component vanishes because of mode coupling between the individual modes at the interface of the upper and lower surfaces. Thus, a Bloch-type field topological defect forms [Fig. 3(c)]. In addition, for waveguide modes [Fig. 2(a)], radial, azimuthal, and normal electric field components contribute simultaneously, and a twisted-type field topological defect forms [Fig. 3(d)]. Similarly, a photonic analog of topological phase transition occurs by constructing using hybrid EM modes from within the light cone (EM mode with subluminal phase velocity) and beyond it (EM mode with superluminal phase velocity).

**Discussions and conclusions:** To summarize, we proposed a SOC based theory to reveal the inherent relationships among the conservation laws, the SOCs, and various configurations of EM topological defects in confined EM fields. The formation of EM spin topological defects is governed by the conserved total angular momentum. Moreover, the topology of EM spin topological defects is protected by the rotational symmetry of EM system from the local disturbance of spin vector. From the optical Dirac equation, we derived the SOC term for a general EM field, in which the real part relates to the gradient of helical density and the imaginary part to the EM transverse spin given by the vorticities associated with the kinetic Abraham–Poynting momentum. This SOC term describes entirely the spin–orbit interactions of light present. By engineering the anisotropic SOC, we constructed various configurations of the EM spin topological defects, specifically the Néel-, twisted-, and Bloch-types of defects. In addition, we showed that a photonic analog of topological phase transition occurs as a variation in phase velocity of the EM mode from superluminal to subluminal. More interestingly, the theory describes quite generally the formation and transitions of various configurations of electric field topological defects. Therefore, our theory provides in a universal manner a general framework for spin and field topological defects and provides an efficient means for understanding topological phenomena in classical wave fields.

**Acknowledgements**

This work was supported, in part, by the Guangdong Major Project of Basic Research grant 2020B0301030009, the National Natural Science Foundation of China grants 12174266, 12047540, U1701661, 61935013, 61427819, and 61622504, the Leadership of Guangdong province program grant 00201505.

**Author contributions**

P. S. conceived the concept, performed the theory and wrote the draft. All authors contributed to the article.

**Competing interests**

The authors declare no competing interests.

**Data availability**

The data that support the plots within this paper and other findings of this study are available from the corresponding author upon reasonable request.

**References:**

[1] J. M. Kosterlitz, "Nobel lecture: Topological defects and phase transitions," *Rev. Mod. Phys.* **89**, 040501 (2017).

[2] A. Ardaševa and A. Doostmohammadi, "Topological defects in biological matter," *Nat. Rev. Phys.* **4**, 354-356 (2022).

[3] S. Mühlbauer, B. Binz, F. Jonietz, C. Pfleiderer, A. Rosch, A. Neubauer, R. Georgii, and P. Böni, "Skyrmion lattice in a chiral magnet," *Science* **323**, 915-919 (2009).

[4] V. Baltz, A. Manchon, M. Tsoi, T. Moriyama, T. Ono, and Y. Tserkovnyak, "Antiferromagnetic spintronics," *Rev. Mod. Phys.* **90**, 015005 (2018).

[5] A. Manchon, J. Železný, I. M. Miron, T. Jungwirth, J. Sinova, A. Thiaville, K. Garello, and P. Gambardella, "Current-induced spin-orbit torques in ferromagnetic and antiferromagnetic systems," *Rev. Mod. Phys.* **91**, 035004 (2019).

[6] N. Nagaosa and Y. Tokura, "Topological properties and dynamics of magnetic skyrmions," *Nat. Nanotechnol.* **8**, 899-911 (2013).

[7] S. L. Zhang, G. van der Laan, W. W. Wang, A. A. Haghighirad, and T. Hesjedal, "Direct observation of twisted surface skyrmions in bulk crystals," *Phys. Rev. Lett.* **120**, 227202 (2018).

[8] T. Bauer, M. Neugebauer, G. Leuchs, and P. Banzer, "Optical polarization Möbius strips and points of purely transverse spin density," *Phys. Rev. Lett.* **117**, 013601 (2016).

[9] H. Kedia, D. Foster, M. R. Dennis, and William T. M. Irvine, "Weaving knotted vector fields with tunable helicity," *Phys. Rev. Lett.* **117**, 274501 (2016).

[10] C. Guo, M. Xiao, Y. Guo, L. Yuan, and S. Fan, "Meron spin textures in momentum space," *Phys. Rev. Lett.* **124**, 106103 (2020).

[11] N. Papasimakis, V. A. Fedotov, V. Savinov, T. A. Raybould, and N. I. Zheludev, "Electromagnetic toroidal excitations in matter and free space," *Nat. Mater.* **15**, 263-271 (2016).

[12] A. Chong, C. Wan, J. Chen, and Q. Zhan, "Generation of spatiotemporal optical vortices with controllable transverse orbital angular momentum," *Nat. Photon.* **14**, 350-354 (2020).

[13] Y. Shen, Y. Hou, N. Papasimakis, and N. I. Zheludev, "Supertoroidal light pulses as electromagnetic skyrmions propagating in free space," *Nat. Commun.* **12**, 5891 (2021).

[14] C. Wan, Q. Cao, J. Chen, A. Chong, and Q. Zhan, "Toroidal vortices of light," *Nat. Photon.* **16**, 519-522 (2022).

[15] A. Forbes, M. de Oliveira, and M. R. Dennis, "Structured light," *Nat. Photon.* **15**, 253-262 (2021).

[16] C. D. Parmee, M. R. Dennis, and J. Ruostekoski, "Optical excitations of Skyrmions, knotted solitons, and defects in atoms," *Commun. Phys.* **5**, 54 (2022).

[17] Y. Shen, B. Yu, H. Wu, C. Li, Z. Zhu, Anatoly V. Zayats, "Topological transformation and free-space transport of photonic hopfions," *Adv. Photon.* 5(1), 015001 (2023)

[18] S. Tsesses, E. Ostrovsky, K. Cohen, B. Gjonaj, N. H. Lindner, and G. Bartal, "Optical skyrmion lattice in evanescent electromagnetic fields," *Science* **361**, 993-996 (2018).

[19] L. Du, A. Yang, A. V. Zayats and X. Yuan, "Deep-subwavelength features of photonic skyrmions in a confined electromagnetic field with orbital angular momentum," *Nat. Phys.* **15**, 650-654 (2019).

[20] T. J. Davis, D. Janoschka, P. Dreher, B. Frank, Frank-J. Meyer zu Heringdorf, and H. Giessen, "Ultrafast vector imaging of plasmonic skyrmion dynamics with deep subwavelength resolution," *Science* **368**, eaba6415 (2020).

[21] S. Tsesses, K. Cohen, E. Ostrovsky, B. Gjonaj, G. Bartal, "Spin–orbit interaction of light in plasmonic lattices," *Nano Lett.* **19**, 4010-4016 (2019).

[22] C. Li, P. Shi, L. Du, and X. Yuan, "Mapping the near-field spin angular momenta in the structured surface plasmon polariton field," *Nanoscale* **12**, 13674-13679 (2020).

[23] P. Shi, L. Du, and X. Yuan, "Spin photonics: from transverse spin to photonic skyrmions," *Nanophotonics* **10**, 3927-3943 (2021).

[24] Y. Dai, Z. Zhou, A. Ghosh, R. S. K. Mong, A. Kubo, C.-B. Huang, and H. Petek, "Plasmonic topological quasiparticle on the nanometre and femtosecond scales," *Nature* **588**, 616-619 (2020).

[25] Y. Dai, Z. Zhou, A. Ghosh, K. Kapoor, M. Dąbrowski, A. Kubo, C.-B. Huang, and H. Petek, "Ultrafast microscopy of a twisted plasmonic spin skyrmion," *Appl. Phys. Rev.* **9**, 011420 (2022).

[26] X. Lei, A. Yang, P. Shi, Z. Xie, L. Du, A. V. Zayats, and X. Yuan, "Photonic spin lattices: symmetry constraints for skyrmion and meron topologies," *Phys. Rev. Lett.* **127**, 237403 (2021).

[27] A. Ghosh, S. Yang, Y. Dai, Z. Zhou, T. Wang, C. Huang, and H. Petek, "A topological lattice of plasmonic merons," *Appl. Phys. Rev.* **8**, 041413 (2021).

[28] P. Shi, X. Lei, Q. Zhang, H. Li, L. Du, and X. Yuan, "Intrinsic spin-momentum dynamics of surface electromagnetic waves in dispersive interfaces," *Phys. Rev. Lett.* **128**, 213904 (2022).

[29] P. Shi, L. Du, M. Li, and X. Yuan, "Symmetry-protected photonic chiral spin textures by spin-orbit coupling," *Laser Photonics Rev.* **15**, 202000554 (2021).

[30] F. J. Rodríguez-Fortuño, N. Engheta, A. Martínez, and A. V. Zayats, "Lateral forces on circularly polarizable particles near a surface," *Nat. Commun.* **6**, 8799 (2015).

[31] S. B. Wang and C. T. Chan, "Lateral optical force on chiral particles near a surface," *Nat. Commun.* **5**, 3307 (2014).

[32] A. Hayat, J. P. B. Mueller, and F. Capasso, "Lateral chirality-sorting optical forces," *Proc. Natl. Acad. Sci. U.S.A.* **112**, 13190-13194 (2015).

[33] Y. Shi, T. Zhu, T. Zhang, A. Mazzulla, D.-P. Tsai, W. Ding, A.-Q. Liu, G. Cipparrone, J. José Sáenz, and C.-W. Qiu, "Chirality-assisted lateral momentum transfer for bidirectional enantioselective separation," *Light Sci. Appl.* **9**, 62 (2020).

[34] M. Gu, Q. Zhang, and S. Lamon, "Nanomaterials for optical data storage," *Nat. Rev. Mater.* **1**, 16070 (2016).

[35] M. Dąbrowski, Y. Dai, and H. Petek, "Ultrafast photoemission electron microscopy: imaging plasmons in space and time," *Chem. Rev.* **120**, 6247-6287 (2020).

[36] F. J. Rodríguez-Fortuño, G. Marino, P. Ginzburg, D. O'Connor, A. Martínez, G. A. Wurtz, and A. V. Zayats, "Near-field interference for the unidirectional excitation of electromagnetic guided modes," *Science* **340**, 328-330 (2013).

[37] J. Petersen, J. Volz, and A. Rauschenbeutel, "Chiral nanophotonic waveguide interface based on spin-orbit interaction of light," *Science* **346**, 67-71 (2014).

[38] P. Lodahl, S. Mahmoodian, S. Stobbe, A. Rauschenbeutel, P. Schneeweiss, J. Volz, H. Pichler, and P. Zoller, "Chiral quantum optics," *Nature* **541**, 473-480 (2017).

[39] M. Mruczkiewicz, M. Krawczyk, and K. Y. Guslienko, "Spin excitation spectrum in a magnetic nanodot with continuous transitions between the vortex, Bloch-type skyrmion, and Néel-type skyrmion states," *Phys. Rev. B* **95**, 094414 (2017).

[40] S. Hayami and Y. Motome, "Néel-and Bloch-type magnetic vortices in Rashba metals," *Phys. Rev. Lett.* **121**, 137202 (2018).

[41] X. Z. Yu, W. Koshibae, Y. Tokunaga, K. Shibata, Y. Taguchi, N. Nagaosa, and Y. Tokura, "Transformation between meron and skyrmion topological spin textures in a chiral magnet," *Nature* **564**, 95-98 (2018).

[42] L. Peng, R. Takagi, W. Koshibae, K. Shibata, K. Nakajima, T. Arima, N. Nagaosa, S. Seki, X. Yu, and Y. Tokura, "Controlled transformation of skyrmions and antiskyrmions in a non-centrosymmetric magnet," *Nat. Nanotechnol.* **15**, 181-186 (2020).

[43] K. Y. Bliokh, F. J. Rodríguez-Fortuño, F. Nori, and A. V. Zayats, "Spin–orbit interactions of light," *Nat. Photon.* **9**, 796-808 (2015).

[44] P. Shi, A. Yang, F. Meng, J. Chen, Y. Zhang, Z. Xie, L. Du, X. Yuan, "Optical near-field measurement for spin-orbit interaction of light," *Prog. in Quantum Electron.* **78**, 100341 (2021).

[45] Q. Zhang, Z. Xie, L. Du, P. Shi, and X. Yuan, "Bloch-type photonic skyrmions in optical chiral multilayers," *Phys. Rev. Research* **3**, 023109 (2021).

[46] Q. Zhang, Z. Xie, P. Shi, H. Yang, H. He, L. Du, and X. Yuan, "Optical topological lattices of Bloch-type skyrmion and meron topologies," *Photon. Res.* **10**, 947-957 (2022).

[47] J. Valentine, S. Zhang, T. Zentgraf, E. Ulin-Avila, D. A. Genov, G. Bartal, and X. Zhang, "Three-dimensional optical metamaterial with a negative refractive index," *Nature* **455**, 376–379 (2008).

[48] W. Gao, M. Lawrence, B. Yang, F. Liu, F. Fang, B. Béri, J. Li, and S. Zhang, "Topological photonic phase in chiral hyperbolic metamaterials," *Phys. Rev. Lett.* **114**, 037402 (2015).

[49] H. Ge, X.-Y. Xu, L. Liu, R. Xu, Z.-K. Lin, S.-Y. Yu, M. Bao, J.-H. Jiang, M.-H. Lu, and Y.-F. Chen, "Observation of acoustic skyrmions," *Phys. Rev. Lett.* **127**, 144502 (2021).

[50] R. D. Muelas-Hurtado, K. Volke-Sepúlveda, J. L. Ealo, F. Nori, M. A. Alonso, K. Y. Bliokh, and E. Brasselet, "Observation of polarization singularities and topological textures in sound waves," *Phys. Rev. Lett.* **129**, 204301 (2022).

[51] L. Cao, S. Wan, Y. Zeng, Y. Zhu, and B. Assouar, "Observation of phononic skyrmions based on hybrid spin of elastic waves," *Sci. Adv.* **9**(7), eadf3652 (2023).

[52] S. M. Barnett, "Optical Dirac equation," *New J. Phys.* **16**, 093008 (2014).

[53] I. Bialynicki-Birula and Z. Bialynicka-Birula, "The role of the Riemann–Silberstein vector in classical and quantum theories of electromagnetism," *J. Phys. A: Math. Theor.* **46**, 053001 (2013).

[54] Shun-Qing Shen, *Topological Insulators: Dirac Equation in Condensed Matters* (Springer-Verlag Berlin Heidelberg, 2012).

[55] M. V. Berry, "Optical currents," *J. Opt. A: Pure Appl. Opt.* **11**, 094001 (2009).

[56] F. Alpeggiani, K. Y. Bliokh, F. Nori, and L. Kuipers, "Electromagnetic helicity in complex media," *Phys. Rev. Lett.* **120**, 243605 (2018).

[57] P. Shi, A. Yang, X. Yin, L. Du, X. Lei, and X. Yuan, Spin decomposition and topological properties in a generic electromagnetic field, *arXiv*:2108.00725 [physics.optics] (2022).

[58] P. Shi, L. Du, C. Li, A. V. Zayats, and X. Yuan, "Transverse spin dynamics in structured electromagnetic guided waves," *Proc. Natl. Acad. Sci. U.S.A.* **118**, e2018816118 (2021).

[59] O. V. Angelsky, A. Y. Bekshaev, S. G. Hanson, C. Yu Zenkova, I. I. Mokhun, and Z. Jun, "Structured light: ideas and concepts," *Front. Phys.* **8**, 114 (2020).

[60] K. Y. Bliokh, D. Smirnova, and F. Nori, "Quantum spin Hall effect of light," *Science* **348**, 1448-1451 (2015).

[61] S. Perea-Puente and F. J. Rodríguez-Fortuño, "Dependence of evanescent wave polarization on the losses of guided optical modes," *Phys. Rev. B* **104**, 085417 (2021).

[62] Allen H. J. Yang, S. D. Moore, B. S. Schmidt, M. Klug, M. Lipson, and D. Erickson, "Optical manipulation of nanoparticles and biomolecules in sub-wavelength slot waveguides," *Nature* **457**, 71-75 (2009).

[63] J. K. Asbóth, L. Oroszlány, and A. Pályi, *A Short Course on Topological Insulators: Band Structure and Edge States in One and Two Dimensions* (Springer Switzerland, 2016).

[64] Z. A. Kudyshev, A. V. Kildishev, A. Boltasseva and V. M. Shalaev, "Photonic topological phase transition on demand," *Nanophotonics* **8**(8), 1349-1356 (2019).

# Supplemental materials for Topological state transitions in electromagnetic topological defects

Peng Shi[1,†], Qiang Zhang[1], and Xiaocong Yuan[2,1‡]

[1]*Nanophotonics Research Centre, Institute of Microscale Optoelectronics & State Key Laboratory of Radio Frequency Heterogeneous Integration, Shenzhen University, Shenzhen 518060, China*
[2]*Zhejiang Lab, Research Center for Humanoid Sensing, Research Institute of Intelligent Sensing, Hangzhou 311100, China*

*Corresponding author:* †pittshiustc@gmail.com; ‡xcyuan@szu.edu.cn

## Contents:

# VI. EM topological defects in C4 and C6 rotating symmetry

# I. Classical Lorentz model of dielectric function

In classical electromagnetic (EM) theory, the wave-matter interactions can be described by the Maxwell's equations [S1]:

$$\begin{aligned}\nabla \cdot \mathbf{D} &= 0 \\ \nabla \times \mathbf{E} &= -\frac{\partial \mathbf{B}}{\partial t} \\ \nabla \cdot \mathbf{B} &= 0 \\ \nabla \times \mathbf{H} &= \mathbf{J}_\sigma + \frac{\partial \mathbf{D}}{\partial t}\end{aligned} \quad , \tag{S1}$$

with the constitutive relations in an isotropic medium expressed as

$$\begin{aligned}\mathbf{D} &= \varepsilon_0 \mathbf{E} + \mathbf{P} \\ \mathbf{B} &= \mu \mathbf{H} \\ \mathbf{J}_\sigma &= \sigma \mathbf{E}\end{aligned} \quad . \tag{S2}$$

Here, $\mathbf{D}$ is electric displacement vector; $\mathbf{E}$ is the electric field strength; $\mathbf{B}$ is the magnetic induction strength; $\mathbf{H}$ is the magnetic field strength; $\mathbf{J}_\sigma$ is the current density; $\varepsilon_0$ is the permittivity of vacuum; $\mathbf{P}$ is the electron polarization; $\mu$ is the magnetic permeability and there is $\mu = \mu_0$ through the manuscript; $\sigma$ is the electron conductance. We do not consider the free charges $\rho_f = 0$.

Then, we consider the Lorentz model of electric permittivity in medium. The electron gas subjected to an external electric field $\mathbf{E}$ can be given by [S1]:

$$m_e \left( \ddot{\mathbf{r}} + \gamma \dot{\mathbf{r}} + \omega_i^2 \mathbf{r} \right) = -e \mathbf{E}(t). \tag{S3}$$

Here, $\mathbf{r}$ is the position vector of electron; $m_e$ is the mass of electron; the electrons oscillate in response to the applied electromagnetic (EM) field $\mathbf{E}(t)$, and their motion is damped with a characteristic frequency $\gamma = 1/\tau$ and $\tau$ is the relaxation time of electron gas. We assume that there are diversiform types of electrons, including the free electron and polarized electrons; $\omega_i$ is the displacement electron resonance frequency of $i$-th type electrons and there is $\omega_i = 0$ for the metal materials. If we assume a harmonic time dependence $\mathbf{E}(t) = \mathbf{E}_0 e^{-i\omega t}$ of the driving field (we use this time convention throughout the manuscript and supplementary materials), a solution of this equation describing the oscillation of the electron is

$$\mathbf{r}_i = \frac{e}{m_e} \frac{1}{\left( \omega^2 - \omega_i^2 + i\omega\gamma \right)} \mathbf{E} , \tag{S4}$$

and the velocity is

$$\mathbf{v}_i = \dot{\mathbf{r}}_i = -\frac{e}{m_e} \frac{i\omega}{\left( \omega^2 - \omega_i^2 + i\omega\gamma \right)} \mathbf{E} . \tag{S5}$$

Firstly, for the metal materials, there are $\mathbf{J}_\sigma \neq 0$ and electric polarization $\mathbf{P} = 0$. Thus, the current density is

$$\mathbf{J}_\sigma = -N\sum_i e\mathbf{v}_i = \frac{Ne^2}{m_e}\sum_i \frac{i\omega f_i}{\left(\omega^2 - \omega_i^2 + i\omega\gamma\right)}\mathbf{E} = \sigma\mathbf{E} \ . \tag{S6}$$

In the special case that there is only one type of free electron, the current can be downgraded into

$$\mathbf{J}_\sigma = -Ne\mathbf{v}_i = \frac{Ne^2}{m_e}\frac{i\omega}{\left(\omega^2 - \omega_i^2 + i\omega\gamma\right)} = \sigma\mathbf{E} \tag{S7}$$

and the fourth equation of Maxwell's equation can be written as

$$\nabla\times\mathbf{H} = \mathbf{J}_\sigma + \frac{\partial\mathbf{D}}{\partial t} = -i\omega\left(\varepsilon_0 + i\frac{\sigma_\omega}{\omega}\right)\mathbf{E} = -i\omega\varepsilon_0\left(1 - \frac{\omega_p^2}{\left(\omega^2 - \omega_i^2 + i\omega\gamma\right)}\right)\mathbf{E} = -i\omega\varepsilon(\omega)\mathbf{E} = \frac{\partial\mathbf{D}(\omega)}{\partial t} \ . \tag{S8}$$

Here, $N$ is the density of free electrons; $\omega_p = \sqrt{Ne^2/m_e\varepsilon_0}$ is the plasma frequency and $\omega_i = 0$.

On the other hand, for the dielectric materials, there are $\mathbf{J}_\sigma = 0$ and electric polarization $\mathbf{P}\neq 0$ ($\mathbf{D} = \varepsilon_0\mathbf{E} + \mathbf{P}$). The electric dipole moment $\mathbf{p}$ is

$$\mathbf{p} = -e\mathbf{r} = -\frac{e^2}{m_e}\frac{1}{\left(\omega^2 - \omega_i^2 + i\omega\gamma\right)}\mathbf{E} \ , \tag{S9}$$

and the electron polarization is

$$\mathbf{P} = -\varepsilon_0\frac{Ne^2}{m_e\varepsilon_0}\sum_i \frac{f_i}{\left(\omega^2 - \omega_i^2 + i\omega\gamma\right)}\mathbf{E} = \varepsilon_0\chi_P\mathbf{E} \ . \tag{S10}$$

In the special case that there is only one type of electric dipole moment, the electron polarization can be downgraded into

$$\mathbf{P} = -\varepsilon_0\frac{Ne^2}{m_e\varepsilon_0}\frac{1}{\left(\omega^2 - \omega_i^2 + i\omega\gamma\right)}\mathbf{E} = \varepsilon_0\chi_P\mathbf{E} \tag{S11}$$

and the fourth equation of Maxwell's equation can be written as

$$\nabla\times\mathbf{H} = \mathbf{J}_\sigma + \frac{\partial\mathbf{D}}{\partial t} = -i\omega\left(\varepsilon_0\mathbf{E} + \mathbf{P}\right) = -i\omega\varepsilon_0\left(1 - \frac{\omega_p^2}{\left(\omega^2 - \omega_i^2 + i\omega\gamma\right)}\right)\mathbf{E} = -i\omega\varepsilon(\omega)\mathbf{E} = \frac{\partial\mathbf{D}(\omega)}{\partial t} \ . \tag{S12}$$

Here, $\omega_p = \sqrt{Ne^2/m_e\varepsilon_0}$ is consistent with the plasma frequency.

Overall, from the equation (S8) and (S12), no matter for the metal or for the dielectric, the permittivity $\varepsilon(\omega)$ can be expressed universally by

$$\varepsilon(\omega) = \varepsilon_0\left(1 - \frac{\omega_p^2}{\left(\omega^2 - \omega_i^2 + i\omega\gamma\right)}\right) . \tag{S13}$$

Therefore, the Maxwell's equations and Helmholtz equations of EM field can be expressed as

$$\nabla \cdot \mathbf{D}(\omega) = 0$$
$$\nabla \times \mathbf{E} = -\frac{\partial \mathbf{B}}{\partial t}$$
$$\nabla \cdot \mathbf{B} = 0 \qquad (S14)$$
$$\nabla \times \mathbf{H} = \frac{\partial \mathbf{D}(\omega)}{\partial t}$$

and

$$\nabla^2 \mathbf{E} + k^2(\omega)\mathbf{E} = 0$$
$$\nabla^2 \mathbf{H} + k^2(\omega)\mathbf{H} = 0, \qquad (S15)$$

respectively. Here, $\varepsilon(\omega)$ is a complex value and depends on the EM frequency. Thus, the wavenumber $k^2(\omega) = \omega^2\varepsilon(\omega)\mu = \omega^2\varepsilon(\omega)\mu_0$ is also complex, where the real part of $k(\omega)$ is related to the propagating property and the imaginary part of $k(\omega)$ is related to the attenuation.

## II. Optical Dirac equation and spin-orbit couplings of EM fields

To study the spin-orbit couplings (SOCs) in EM field, it needs to introduce the optical Dirac equation. The Dirac equation is a relativistic quantum mechanical one for elementary spin-1/2 particle [54], whereas with the identity for arbitrary two vectors **A** and **B**: $\mathbf{A} \times \mathbf{B} = -i(\mathbf{A} \cdot \hat{\mathbf{S}})\mathbf{B}$ [55], where $\hat{\mathbf{S}}$ is the spin-1 matrix in SO(3) expressed as:

$$\hat{\mathbf{S}} = \{\hat{S}_x, \hat{S}_y, \hat{S}_z\} = \left\{ \begin{pmatrix} 0 & 0 & 0 \\ 0 & 0 & i \\ 0 & -i & 0 \end{pmatrix}, \begin{pmatrix} 0 & 0 & -i \\ 0 & 0 & 0 \\ i & 0 & 0 \end{pmatrix}, \begin{pmatrix} 0 & i & 0 \\ -i & 0 & 0 \\ 0 & 0 & 0 \end{pmatrix} \right\}, \qquad (S16)$$

the Maxwell's equations in the form of the Dirac equation named as optical Dirac equation have been introduced and researched for years [52]

$$\hat{\mathbf{H}}|\Psi\rangle = v \begin{pmatrix} \mathbf{0} & \hat{\mathbf{S}} \\ \hat{\mathbf{S}} & \mathbf{0} \end{pmatrix} \cdot \hat{\mathbf{p}}|\Psi\rangle = v\hat{\boldsymbol{\alpha}} \cdot \hat{\mathbf{p}}|\Psi\rangle = i\hbar\frac{\partial}{\partial t}|\Psi\rangle, \qquad (S17)$$

with the Riemann–Silberstein (RS) vector (6-vector photon wave function) given by [53]

$$|\Psi\rangle = \frac{1}{2}\begin{pmatrix} \sqrt{\varepsilon}\mathbf{E} \\ i\sqrt{\mu}\mathbf{H} \end{pmatrix}. \qquad (S18)$$

Here, $v = 1/\sqrt{\varepsilon\mu}$ is the velocity of light in medium. In the form of RS vector, the Minkowski-type canonical momentum and spin angular momentum (SAM) can be expressed as

$$\mathbf{P}^M = \frac{1}{4\omega}\text{Im}\{\varepsilon\mathbf{E}^* \cdot (\nabla)\mathbf{E} + \mu\mathbf{H}^* \cdot (\nabla)\mathbf{H}\} = \frac{1}{\hbar\omega}\langle\Psi|\begin{bmatrix} \hat{\mathbf{p}} & \mathbf{0} \\ \mathbf{0} & \hat{\mathbf{p}} \end{bmatrix}|\Psi\rangle = \frac{1}{\hbar\omega}\langle\Psi|\hat{\mathbf{P}}|\Psi\rangle, \qquad (S19)$$

and

$$\boldsymbol{\Sigma} = \frac{1}{4\omega}\text{Im}\{\varepsilon\mathbf{E}^* \times \mathbf{E} + \mu\mathbf{H}^* \times \mathbf{H}\} = \frac{1}{\hbar\omega}\langle\Psi|\hbar\begin{bmatrix} \hat{\mathbf{S}} & \mathbf{0} \\ \mathbf{0} & \hat{\mathbf{S}} \end{bmatrix}|\Psi\rangle = \frac{1}{\hbar\omega}\langle\Psi|\hat{\boldsymbol{\Sigma}}|\Psi\rangle. \qquad (S20)$$

Here, $\hat{\mathbf{p}} = -i\hbar\nabla$ is the momentum operator. Therefore, the orbital angular momentum (OAM) operator and intrinsic SAM operator are

$$\hat{\mathbf{L}} = \hat{\mathbf{r}} \times \hat{\mathbf{p}} \tag{S21}$$

and

$$\hat{\mathbf{\Sigma}} = \hbar \begin{bmatrix} \hat{\mathbf{S}} & 0 \\ 0 & \hat{\mathbf{S}} \end{bmatrix}, \tag{S22}$$

respectively.

Firstly, the time derivative of position operator $\hat{\mathbf{r}}$ can be calculated as

$$\dot{\hat{\mathbf{r}}} = \frac{i}{\hbar}\left[\hat{\mathbf{H}}, \hat{\mathbf{r}}\right] = v \begin{bmatrix} 0 & \hat{\mathbf{S}} \\ \hat{\mathbf{S}} & 0 \end{bmatrix} = v\hat{\boldsymbol{\alpha}}. \tag{S23}$$

Thus, the underlying physics of the operator $v\hat{\boldsymbol{\alpha}}$ can be understood as the group velocity of EM field which given by the Poynting vector because there is

$$\mathbf{p}^A = \frac{1}{2}\mathrm{Re}\{\mathbf{E}^* \times \mathbf{H}\} = \frac{1}{2}\mathrm{Re}\{-i(\mathbf{E}^* \cdot \hat{\mathbf{S}})\mathbf{H}\} = \langle\Psi|v\begin{bmatrix} 0 & \hat{\mathbf{S}} \\ \hat{\mathbf{S}} & 0 \end{bmatrix}|\Psi\rangle = \langle\Psi|v\hat{\boldsymbol{\alpha}}|\Psi\rangle. \tag{S24}$$

Notable that the Poynting vector $\mathbf{p}^A$ determines the kinetic Abraham-Poynting momentum $\mathbf{P}^A$ in the EM system $\mathbf{P}^A \propto \mathbf{p}^A/v^2$ [56].

Secondly, the conserved properties of SAM and OAM can be expressed as

$$\dot{\hat{\mathbf{\Sigma}}} = \frac{i}{\hbar}\left[\hat{\mathbf{H}}, \hat{\mathbf{\Sigma}}\right] = -v\hat{\boldsymbol{\alpha}} \times \hat{\mathbf{p}} \text{ and } \dot{\hat{\mathbf{L}}} = \frac{i}{\hbar}\left[\hat{\mathbf{H}}, \hat{\mathbf{L}}\right] = v\hat{\boldsymbol{\alpha}} \times \hat{\mathbf{p}}. \tag{S25}$$

These equations definitely show that the SAM and OAM are not conserved separately in a general EM system. However, the total angular momentum (AM) operator $\hat{\mathbf{J}} = \hat{\mathbf{L}} + \hat{\mathbf{\Sigma}}$ is conserved owing to

$$\dot{\hat{\mathbf{J}}} = \frac{i}{\hbar}\left[\hat{\mathbf{H}}, \hat{\mathbf{J}}\right] = \frac{i}{\hbar}\left[\hat{\mathbf{H}}, \hat{\mathbf{L}}\right] + \frac{i}{\hbar}\left[\hat{\mathbf{H}}, \hat{\mathbf{\Sigma}}\right] = 0. \tag{S26}$$

Obviously, the SOC is related to the operator $\hat{\mathbf{H}}_{\mathrm{SO}} = v\hat{\boldsymbol{\alpha}} \times \hat{\mathbf{p}}$. In this way, the spin-orbit interactions can be calculated as

$$\mathbf{H}_{\mathrm{SO}} = \langle\Psi|\hat{\mathbf{H}}_{\mathrm{SO}}|\Psi\rangle = \langle\Psi|v\begin{bmatrix} 0 & \hat{\mathbf{S}} \times \hat{\mathbf{p}} \\ \hat{\mathbf{S}} \times \hat{\mathbf{p}} & 0 \end{bmatrix}|\Psi\rangle = \frac{i\hbar}{4}\left[(\mathbf{H}^* \cdot \nabla)\mathbf{E} - (\mathbf{E}^* \cdot \nabla)\mathbf{H}\right]. \tag{S27}$$

Therein, the real part of $\mathbf{H}_{\mathrm{SO}}$ is

$$\mathrm{Re}\,\mathbf{H}_{\mathrm{SO}} = \mathrm{Re}\left\{\frac{i\hbar}{4}\left[(\mathbf{H}^* \cdot \nabla)\mathbf{E} - (\mathbf{E}^* \cdot \nabla)\mathbf{H}\right]\right\} = \frac{i\hbar}{4}\left[\nabla(\mathbf{H}^* \cdot \mathbf{E} - \mathbf{E}^* \cdot \mathbf{H})\right] = -\frac{\hbar\omega}{k^2}\nabla C. \tag{S28}$$

This term indicates that the SOC is related to the gradient of EM helical density [56]

$$C = -\frac{\varepsilon\omega}{2}\mathrm{Im}\{\mathbf{E}^* \cdot \mathbf{B}\} \tag{S29}$$

in the EM system.

On the other hand, the imaginary part of $\mathbf{H}_{\mathrm{SO}}$ is

$$\mathrm{Im}\,\mathbf{H}_{\mathrm{SO}} = \mathrm{Im}\left\{\frac{i\hbar}{4}\left[(\mathbf{H}^* \cdot \nabla)\mathbf{E} - (\mathbf{E}^* \cdot \nabla)\mathbf{H}\right]\right\} = \nabla \times \frac{\hbar}{4}\mathrm{Re}\{\mathbf{E} \times \mathbf{H}^*\} = \frac{\hbar}{2}\nabla \times \mathbf{p}^A. \tag{S30}$$

This term indicates that the SOC is also related to the curl of Poynting vector/kinetic Abraham-Poynting momentum, which is proportional to the EM transverse spin

$$\mathbf{S}_T = \frac{1}{2\omega^2}\nabla\times\mathbf{p}^A = \frac{1}{2k^2}\nabla\times\mathbf{P}^A \qquad (S31)$$

in the nondispersive medium [56]. This quantity of SOC in Eq. (S25), whose real part and imaginary part are given in Eq. (S28) and Eq. (S30), respectively, is one of key achievement in our manuscript.

To understand these SOC term, we take the paraxial Hermite-Gaussian (HG) optical beam propagating in the $z$-direction in Cartesian coordinates $(x, y, z)$, in which no spin-orbit AM conversions exist approximatively, for example. However, there will still be SOCs in the system. The electric and magnetic field components can be expressed as [S2]

$$\mathbf{E}_{\mathrm{HG}} = \left[+\eta_x u_{\mathrm{HG}}\hat{\mathbf{x}}, +\eta_y u_{\mathrm{HG}}\hat{\mathbf{y}}, +\frac{1}{ik}\left(\eta_x\frac{\partial}{\partial x}+\eta_y\frac{\partial}{\partial y}\right)u_{\mathrm{HG}}\hat{\mathbf{z}}\right]^{\mathrm{T}} e^{-ikz} \qquad (S32)$$

and

$$\mathbf{H}_{\mathrm{HG}} = \frac{k}{\omega\mu}\left[+\eta_y u_{\mathrm{HG}}\hat{\mathbf{x}}, -\eta_x u_{\mathrm{HG}}\hat{\mathbf{y}}, +\frac{1}{ik}\left(\eta_y\frac{\partial}{\partial x}-\eta_x\frac{\partial}{\partial y}\right)u_{\mathrm{HG}}\hat{\mathbf{z}}\right]^{\mathrm{T}} e^{-ikz}, \qquad (S33)$$

where $\eta_x$ and $\eta_y$ are arbitrary complex constants describing the relative strength, $\mathrm{Im}\{\eta_x^*\eta_y\}$ specifies the polarization ellipticity (helicity) of the paraxial HG beam, and the superscript T indicates the transpose of the matrix. The complex amplitude $u_{\mathrm{HG}}$ is given by

$$u_{\mathrm{HG},mn} = \frac{w_0}{w(z)}H_m\left[\frac{\sqrt{2}x}{w(z)}\right]H_n\left[\frac{\sqrt{2}y}{w(z)}\right]\exp\left(-\frac{x^2+y^2}{w^2(z)}-i\frac{k(x^2+y^2)}{2R(z)}\right)\exp\left(-i(1+m+n)\tan^{-1}\left(\frac{z}{z_R}\right)\right). \qquad (S34)$$

Here, $H_m(x)$ is the Hermite polynomial with non-negative integer index $m$, $z_R = \pi w_0^2/\lambda$ the Rayleigh range, $w(z) = w_0\sqrt{1-z^2/z_R^2}$ the beam width of the propagating wave, $w_0$ the beam radius at the beam waist, $R(z)$ the radius of curvature of the wavefronts, $\lambda$ the wavelength, and the last factor $\exp(-i(1+m+n)\tan^{-1}(z/z_R))$ is the Gouy phase.

The EM helicity of paraxial HG beam and the gradient of EM helicity are

$$C_{\mathrm{HG}} = \frac{\varepsilon k}{2\mu}\mathrm{Im}\left[\eta_y^*\eta_x - \eta_y\eta_x^*\right]u_{\mathrm{HG}}^*u_{\mathrm{HG}} \qquad (S35)$$

and

$$\nabla C_{\mathrm{HG}} = \frac{\varepsilon k}{2\mu}\mathrm{Im}\left[\eta_y^*\eta_x - \eta_y\eta_x^*\right]\left[\frac{\partial u_{\mathrm{HG}}^*u_{\mathrm{HG}}}{\partial x}\hat{\mathbf{x}}, \frac{\partial u_{\mathrm{HG}}^*u_{\mathrm{HG}}}{\partial y}\hat{\mathbf{y}}, 0\hat{\mathbf{z}}\right]^{\mathrm{T}}, \qquad (S36)$$

respectively. Here, the $z$-component is ignored in the paraxial approximation ($\frac{\partial^2 u_{\mathrm{HG}}}{\partial z^2} \ll k\frac{\partial u_{\mathrm{HG}}}{\partial z} \ll k^2 u_{\mathrm{HG}}$). The horizontal components of this gradient of EM helicity are similar to the projections of the longitudinal spin component ($\mathbf{S}_L$) of paraxial HG beam into the $xy$-plane

$$\mathbf{S}_L = \frac{\varepsilon}{4\omega} \text{Im} \left\{ \begin{array}{l} \frac{1}{ik}\left[-\left(\eta_x^*\eta_y - \eta_x\eta_y^*\right)\left(u_{HG}^* \frac{\partial u_{HG}}{\partial x} - u_{HG}\frac{\partial u_{HG}^*}{\partial x}\right)\right]\hat{\mathbf{x}} \\ \frac{1}{ik}\left[-\left(\eta_x^*\eta_y - \eta_x\eta_y^*\right)\left(u_{HG}^* \frac{\partial u_{HG}}{\partial y} - u_{HG}\frac{\partial u_{HG}^*}{\partial y}\right)\right]\hat{\mathbf{y}} \\ \left[\begin{array}{l} 2\left(\eta_x^*\eta_y - \eta_y^*\eta_x\right)u_{HG}^* u_{HG} \\ -\frac{1}{k^2}\left(\eta_x^*\eta_x + \eta_y^*\eta_y\right)\left(\nabla u_{HG}^* \times \nabla u_{HG}\right)_z + \frac{1}{2k^2}\left(\eta_x^*\eta_y - \eta_y^*\eta_x\right)\nabla_\perp^2\left(u_{HG}^* u_{HG}\right) \end{array}\right]\hat{\mathbf{z}} \end{array} \right\}. \quad (S37)$$

This term indicates the SOC between the longitudinal spin (EM helicity) and the intrinsic/extrinsic OAMs [43,44], which have led to the spin-orbit interaction phenomena such as spin Hall effect and spin-to-orbit AM conversions, etc. On the other hand, the imaginary part of $\mathbf{H}_{SO}$ given by the vorticities of kinetic Abraham-Poynting momentum is related to the EM transverse spin

$$\mathbf{S}_T = \frac{1}{2k^2}\nabla \times \mathbf{p} = \frac{\varepsilon}{4\omega}\text{Im}\left\{ \begin{array}{l} \frac{1}{ik}\left[+\left(\eta_x^*\eta_x + \eta_y^*\eta_y\right)\frac{\partial u_{HG}^* u_{HG}}{\partial y}\right]\hat{\mathbf{x}} \\ \frac{1}{ik}\left[-\left(\eta_x\eta_x^* + \eta_y\eta_y^*\right)\frac{\partial u_{HG}^* u_{HG}}{\partial x}\right]\hat{\mathbf{y}} \\ \frac{1}{k^2}\left[\left(\eta_x^*\eta_x + \eta_y^*\eta_y\right)\left(\nabla u_{HG}^* \times \nabla u_{HG}\right)_z - \frac{1}{2}\left(\eta_x^*\eta_y - \eta_y^*\eta_x\right)\nabla_\perp^2\left(u_{HG}^* u_{HG}\right)\right]\hat{\mathbf{z}} \end{array} \right\}, \quad (S37)$$

in which the *x*- and *y*- components are related to the inhomogeneities/structural properties of EM field and the *z*-component is related to the Berry curvature of EM system [57]. This term indicates the SOC between the transverse spin and the intrinsic/extrinsic OAMs, which have led to the phenomena such as spin-momentum locking and orbit-to-spin AM conversions [23,58], etc. Notable that the total SAM is given by $\mathbf{S} = \mathbf{S}_L + \mathbf{S}_T$.

In sum, from the former analysis, the term $\mathbf{H}_{SO}$ definitely describes the SOC in the EM system, and we will uncover the mechanisms of the formation of various EM spin topological defects by using this term.

## III. Phenomenology of EM spin skyrmions at the multilayered systems

The phenomenological theory of the formation of specific Néel-type configuration photonic skyrmion at metal surface was performed in Ref. [29]. Here, we aim to generalize the theory to uncover the formation of various configurations of EM spin skyrmions at the multilayered system. The square modulus of total angular momentum can be expressed as

$$\mathbf{J}\cdot\mathbf{J} = \mathbf{L}\cdot\mathbf{L} + 2\mathbf{L}\cdot\mathbf{S} + \mathbf{S}\cdot\mathbf{S}. \quad (S38)$$

In the multilayered system, there is cylindrical symmetry, which is associated with the conservation of the square modulus of total AM in the normal direction (assuming in the *z*-direction in the cylindrical coordinates ($r$, $\varphi$, $z$)). Thus, the OAM and SAM components of Eq. (S38) are $\mathbf{L} = \mathbf{r} \times \mathbf{P}^M$ and $\mathbf{S} = \mathbf{r} \times \mathbf{P}^S$, respectively. Therein, the kinetic Abraham-Poynting momentum is decomposed into the Minkowski-type canonical momentum $\mathbf{P}^M$ and the Belinfante spin momentum $\mathbf{P}^S$, and the Belinfante spin momentum is given by $\mathbf{P}^S = \nabla \times \mathbf{\Sigma}/2$. Obviously, in physics, the variation of the integration of $\mathbf{J}\cdot\mathbf{J}$ on the local spin vector in the whole plane perpendicular to *z*-axis should be

zero:

$$\delta \langle \mathbf{J} \cdot \mathbf{J} \rangle = \delta \int_\infty \mathbf{J} \cdot \mathbf{J} d\Omega = \delta \int_\infty \mathbf{L} \cdot \mathbf{L} d\Omega + \delta \int_\infty 2\mathbf{L} \cdot \mathbf{S} d\Omega + \delta \int_\infty \mathbf{S} \cdot \mathbf{S} d\Omega$$
$$= \delta \int_\infty \mathbf{r} \cdot \mathbf{r} P_\varphi^{M2} d\Omega + \delta \int_\infty 2\mathbf{r} \cdot \mathbf{r} P_\varphi^M P_\varphi^S d\Omega + \delta \int_\infty \mathbf{r} \cdot \mathbf{r} P_\varphi^{S2} d\Omega \quad . \tag{S39}$$

Because we only interest in the geometry of the directional vector of spin texture, we set **r** as the unit directional vector and one has **r**·**r** =1, and then the Eq. (S39) can be translated into

$$\delta \langle \mathbf{J} \cdot \mathbf{J} \rangle = \delta \int_\infty \left( P_\varphi^{M2} + 2 P_\varphi^M P_\varphi^S + P_\varphi^{S2} \right) d\Omega . \tag{S40}$$

Subsequently, from relativity, the Minkowski-type canonical momentum should be proportional to energy density $W$ and decay by $1/r$-dependent for cylindrical waves (such as the surface Bessel modes [29]):

$$\mathbf{P}^M = \frac{\hbar \ell}{\hbar \omega} \frac{W}{r} . \tag{S41}$$

Here, $\ell$ is the quantum number of OAM. By carefully choosing the relative complex amplitude of electric/magnetic fields, we can obtain $W = \hbar\omega$ for a single wave packet. Therefore, the OAM term is a constant quantity and the variation of OAM term on the local spin vector is zero: $\delta\mathbf{L} = 0$. Then, as the situation of magnetic skyrmions [6], one can assume that the normalized local spin vector $\mathbf{\Sigma}$ has an expression of

$$\mathbf{\Sigma} = \hbar \sigma_h \left[ \sin\Theta \cos\Phi \hat{\mathbf{r}}, \sin\Theta \sin\Phi \hat{\boldsymbol{\varphi}}, \cos\Theta \hat{\mathbf{z}} \right] \tag{S42}$$

by ignoring the dependent of $\Theta$ and $\Phi$ on $z$-axis and considering $\Theta = \Theta(r,\varphi)$ and $\Phi = \Phi(r,\varphi)$ for the spin vector of a single wave packet. In particular, for the Néel-type, Bloch-type and twisted type skyrmions discussed in our manuscript, we can further set that $\Theta = \Theta(r)$ and $\Phi$ = Constant. Then, the variational equation Eq. (S40) can be calculated as

$$\delta \langle \mathbf{J} \cdot \mathbf{J} \rangle = \frac{\hbar^2 \sigma^2}{4} \int_0^\infty \delta\Theta \left\{ \begin{array}{l} -2\left(\sin^2\Theta + \cos^2\Theta \sin^2\Phi\right)\dfrac{\partial^2\Theta}{\partial r^2} \\ -\sin 2\Theta \left(\sin\Phi + \cos 2\Phi\right)\left(\dfrac{\partial\Theta}{\partial r}\right)^2 + \dfrac{2\sin 2\Theta \sin^2\Phi}{r^2} \end{array} \right\} dr , \tag{S43}$$

where we use the calculus

$$\int_0^\infty \left[ 2\left(\sin^2\Theta + \cos^2\Theta \sin^2\Phi\right)\frac{\partial\Theta}{\partial r} + \frac{\sin 2\Theta \sin^2\Phi}{r} \right] \delta\left(\frac{\partial\Theta}{\partial r}\right) dr$$
$$= \int_0^\infty \delta\Theta \left[ \frac{\sin 2\Theta \sin^2\Phi}{r^2} - \frac{2\cos 2\Theta \sin^2\Phi}{r}\frac{\partial\Theta}{\partial r} - 2\sin 2\Theta \cos^2\Phi \left(\frac{\partial\Theta}{\partial r}\right)^2 - 2\left(\sin^2\Theta + \cos^2\Theta \sin^2\Phi\right)\frac{\partial^2\Theta}{\partial r^2} \right] dr . \tag{S44}$$

For the **Néel-type skyrmion**, there is $\Phi = 0$, and the variational equation is downgraded to

$$-2\sin^2\Theta \frac{\partial^2\Theta}{\partial r^2} - \sin 2\Theta \left(\frac{\partial\Theta}{\partial r}\right)^2 = 0 . \tag{S45}$$

The nontrivial solution of this partial differential equation (S45) is

$$\Theta(r) = \pm \arccos\left[-c_1 r - c_1 c_2\right], \tag{S46}$$

The constants $c_1$ and $c_2$ are determined by the boundary conditions.

For the **Bloch-type skyrmion**, there is $\Phi = \pi/2$, and the variational equation is downgraded to

$$-2\frac{\partial^2 \Theta}{\partial r^2} + \frac{2\sin 2\Theta}{r^2} = 0. \tag{S47}$$

By specifying $r = e^t$, the equation is translated to

$$\frac{\partial^2 \Theta}{\partial t^2} - \frac{\partial \Theta}{\partial t} - \sin 2\Theta = 0. \tag{S48}$$

For the **twisted skyrmion**, we can specify that $\Phi = \pi/4$, and the variational equation is downgraded to

$$-(1+\sin^2 \Theta)\frac{\partial^2 \Theta}{\partial r^2} - \frac{\sqrt{2}}{2}\sin 2\Theta \left(\frac{\partial \Theta}{\partial r}\right)^2 + \frac{\sin 2\Theta}{r^2} = 0. \tag{S49}$$

By specifying $r = e^t$, the equation is translated to

$$\frac{\partial^2 \Theta}{\partial t^2} + \frac{\sqrt{2}}{2}\frac{\sin 2\Theta}{1+\sin^2 \Theta}\left(\frac{\partial \Theta}{\partial t}\right)^2 - \frac{\partial \Theta}{\partial t} - \frac{\sin 2\Theta}{1+\sin^2 \Theta} = 0. \tag{S50}$$

The Eq. (S48) and Eq. (S50) can be solved numerically. By solving these partial differential equations by MATHEMATICA, one can obtain the similar curves as that shown in Fig. S1, which indicates the spin vector varies from centre 'up' state to boundary 'down' state gradually without discontinuities/singularities. The skyrmion number of these spin textures

$$N_{SK} = \frac{1}{4\pi}\iint_\Omega \mathbf{M}\cdot\left(\frac{\partial \mathbf{M}}{\partial x}\times\frac{\partial \mathbf{M}}{\partial y}\right)dxdy \tag{S51}$$

with the normalized spin vector $\mathbf{M} = \mathbf{\Sigma}/|\mathbf{\Sigma}|$ is equal to −1 universally and the configurations of EM spin skyrmions can be tuned by the SOCs.

In the section V, we will investigate the SOCs of various EM modes akin to those in magnetic skyrmions to analyze the transitions between the various configurations of EM topological textures.

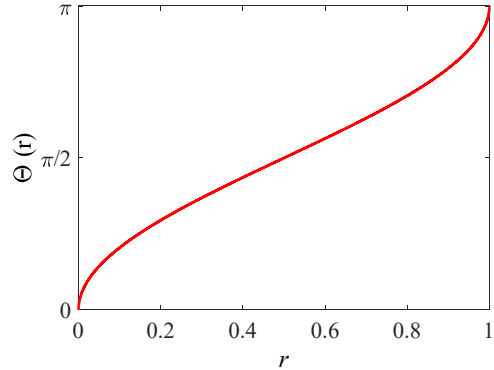

**Fig. S1.** The solution of partial differential equation (S45). The boundary conditions include $\Theta(0) = 0$ and $\Theta(1) = \pi$. For the Eq. (S48) and Eq. (S50), the abscissa axis is $t$.

## IV. Various EM modes and EM field topological structures

### 1. Electric field topological quasiparticles in surface plasmon polariton modes

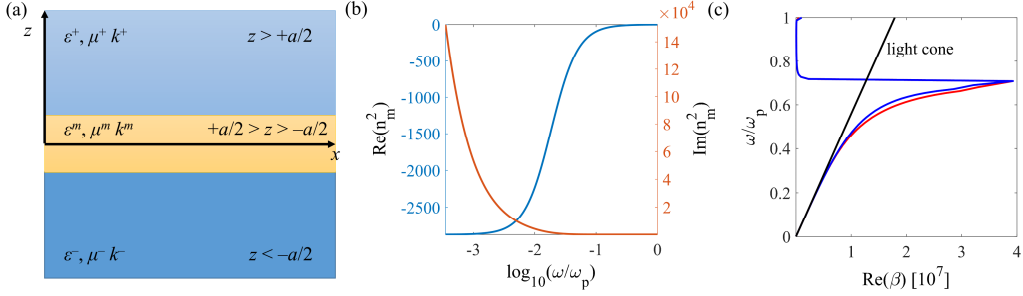

**Fig. S2.** (a) Schematic diagram of one-layer configuration containing lossy metal to excite the *p*-polarized surface modes (surface plasmons polaritons: SPPs). The interfaces are localized between the planes $z = +a/2$ and $z = -a/2$. (b) The permittivity of metal film and (c) the angular frequency ($\omega/\omega_p$) via the propagating constant Re($\beta$). $\omega_p \approx 5.36 \times 10^{15}/s$ ($\lambda_p \approx 0.35 \times 10^{-6}m$) is the plasma frequency and the characteristic frequency $\gamma = 10^{14}/s$. The red/blue lines denote the symmetric/anti-symmetric surface modes of air-metal-air structure with the thickness equal to $0.2\lambda$. The black line represents the light cone of vacuum. $k_0 = \omega/c$ is the wavenumber in vacuum.

Multilayered configuration is beneficial for designing and fabricating the dispersion-engineered artificial metamaterials [47,48]. For the lossy metal in the multilayered systems (**Fig. S2(a)**), the permittivity $\varepsilon(\omega)$ (refractive index $n(\omega)$) is

$$\varepsilon(\omega) = \varepsilon_0 n^2(\omega) = \varepsilon_0 \left(1 - \frac{\omega_p^2}{\omega^2 + i\omega\gamma}\right) = \varepsilon_0 \left(1 - \frac{\omega_p^2}{\omega^2 + \gamma^2} + i\frac{\omega_p^2}{\omega^3/\gamma + \omega\gamma}\right). \tag{S52}$$

For the noble metal, it is naturally that the characteristic frequency $\gamma \approx 10^{14} H_z$ and $\omega_p \approx 10^{16} H_z$. Thus, we use the characteristic frequency $\gamma \approx 10^{14} H_z$ and consider $\omega < \omega_p$. Thus, the real part of permittivity is negative: Re$\{\varepsilon(\omega)\} < 0$ (**Fig. S2(b)**).

For the insulator-metal-insulator system shown in **Fig. S2(a)**, only the *p*-polarized (transverse magnetic, TM) surface EM mode can be excited [S3]. The electric/magnetic field components of the *p*-polarized surface modes in Cartesian coordinates ($x$, $y$, $z$) are summarized in **Table S1**:

**Table S1.** Electric/Magnetic field components of the *p*-polarized surface EM modes

| Region | $z > +\frac{a}{2}$ | $-\frac{a}{2} < z < +\frac{a}{2}$ | $z < -\frac{a}{2}$ |
|---|---|---|---|
| Electric field | $E_x^+ = -\frac{k_z^+}{\beta^2}\frac{\partial E_z^+}{\partial x}$ | $E_x^m = E_x^{m+} + E_x^{m-} = +\frac{k_z^m}{\beta^2}\frac{\partial E_z^{m+}}{\partial x} - \frac{k_z^m}{\beta^2}\frac{\partial E_z^{m-}}{\partial x}$ | $E_x^- = +\frac{k_z^-}{\beta^2}\frac{\partial E_z^-}{\partial x}$ |
| | $E_y^+ = -\frac{k_z^+}{\beta^2}\frac{\partial E_z^+}{\partial y}$ | $E_y^m = E_y^{m+} + E_y^{m-} = +\frac{k_z^m}{\beta^2}\frac{\partial E_z^{m+}}{\partial y} - \frac{k_z^m}{\beta^2}\frac{\partial E_z^{m-}}{\partial y}$ | $E_y^- = +\frac{k_z^-}{\beta^2}\frac{\partial E_z^-}{\partial y}$ |
| | $E_z^+ = \frac{A_+}{\varepsilon^+}\xi e^{-k_z^+(z-a/2)}$ | $E_z^m = E_z^{m+} + E_z^{m-} = \frac{B_+}{\varepsilon^m}\xi e^{+k_z^m(z-a/2)} + \frac{B_-}{\varepsilon^m}\xi e^{-k_z^m(z+a/2)}$ | $E_z^- = \frac{A_-}{\varepsilon^-}\xi e^{+k_z^-(z+a/2)}$ |
| Magnetic field | $H_x^+ = -\frac{i\omega\varepsilon^+}{\beta^2}\frac{\partial E_z^+}{\partial y}$ | $H_x^m = H_x^{m+} + H_x^{m-} = -\frac{i\omega\varepsilon^m}{\beta^2}\frac{\partial E_z^{m+}}{\partial y} - \frac{i\omega\varepsilon^m}{\beta^2}\frac{\partial E_z^{m-}}{\partial y}$ | $H_x^- = -\frac{i\omega\varepsilon^-}{\beta^2}\frac{\partial E_z^-}{\partial y}$ |
| | $H_y^+ = +\frac{i\omega\varepsilon^+}{\beta^2}\frac{\partial E_z^+}{\partial x}$ | $H_y^m = H_y^{m+} + H_y^{m-} = +\frac{i\omega\varepsilon^m}{\beta^2}\frac{\partial E_z^{m+}}{\partial x} + \frac{i\omega\varepsilon^m}{\beta^2}\frac{\partial E_z^{m-}}{\partial x}$ | $H_y^- = +\frac{i\omega\varepsilon^-}{\beta^2}\frac{\partial E_z^-}{\partial x}$ |
| | $H_z^+ = 0$ | $H_z^m = 0$ | $H_z^- = 0$ |

Here, $i = +, -$ and $m$ are corresponding to the regions $z > +a/2$, $z < -a/2$ and $-a/2 < z < +a/2$, respectively. $\xi(x,y)$ is the function of horizontal coordinates (*x,y*) and satisfies the transverse Helmholtz equation $\nabla_\perp^2 \xi + \beta^2 \xi = 0$ with $\nabla_\perp^2 = \partial^2/\partial x^2 + \partial^2/\partial y^2$. By considering the EM boundary conditions, the dispersion relation can be expressed as [S3]

$$e^{-2k_z^m a} = \frac{\left(k_z^m/\varepsilon^m + k_z^+/\varepsilon^+\right)\left(k_z^m/\varepsilon^m + k_z^-/\varepsilon^-\right)}{\left(k_z^m/\varepsilon^m - k_z^+/\varepsilon^+\right)\left(k_z^m/\varepsilon^m - k_z^-/\varepsilon^-\right)}. \tag{S53}$$

Here, the propagating constant (horizontal wavevector) $\beta$ is expressed as

$$\beta^2 = \mathbf{k}^i(\omega) \cdot \mathbf{k}^i(\omega) + k_z^i \cdot k_z^i = \omega^2 \varepsilon^i(\omega) \mu^i + k_z^i \cdot k_z^i, \tag{S54}$$

where $k^i$ is the total wavenumber and $k_z^i$ is z-component wavenumber of *i*-th layer. Here, it is worth noting that the propagating constant $\beta$ is a complex number for the lossy mode [61], where the real part indicates the propagating property and the imaginary part represents the attenuation. The field parameters are

$$\frac{B_+}{B_-} = \frac{k_z^m/\varepsilon^m - k_z^+/\varepsilon^+}{k_z^m/\varepsilon^m + k_z^+/\varepsilon^+} e^{-k_z^m a} \quad \frac{B_-}{B_+} = \frac{k_z^m/\varepsilon^m - k_z^-/\varepsilon^-}{k_z^m/\varepsilon^m + k_z^-/\varepsilon^-} e^{-k_z^m a} \quad \begin{aligned} +A_+ &= +B_+ + B_- e^{-k_z^m a} \\ +B_+ e^{-k_z^m a} + B_- &= +A_- \end{aligned}. \tag{S55}$$

Since only the relative amplitude makes physical sense, one can set one of $A_+$, $B_+$, $B_-$ and $A_-$ to be 1 and the other amplitude coefficients can be calculated properly.

Particularly, for the air-metal-air structure considered in our manuscript, there is $\varepsilon^+ = \varepsilon^- = \varepsilon_0$. Here, $\varepsilon_0$ is the permittivity of air. Therefore, the dispersion relation (S53) can be re-expressed as

$$+e^{-k_z^m a} = \frac{k_z^m/\varepsilon^m + k_z^+/\varepsilon_0}{k_z^m/\varepsilon^m - k_z^+/\varepsilon_0} \tag{S56}$$

for the symmetric modes and

$$-e^{-k_z^m a} = \frac{k_z^m/\varepsilon^m + k_z^+/\varepsilon_0}{k_z^m/\varepsilon^m - k_z^+/\varepsilon_0} \tag{S57}$$

for the anti-symmetric mode. The dispersion relations of the symmetric and antisymmetric modes are shown in the red and blue lines of **Fig. S2(c)**, respectively. If the thickness of layer is thick enough, the left terms of both Eq. (S56) and Eq. (S57) are zero approximatively, and hence the propagation constants can be expressed universally as

$$\beta = \omega \sqrt{\frac{\varepsilon^m \varepsilon_0}{\varepsilon_0 + \varepsilon^m} \mu_0} = \frac{\omega}{c} \sqrt{\frac{\varepsilon^m/\varepsilon_0}{1 + \varepsilon^m/\varepsilon_0}}. \tag{S58}$$

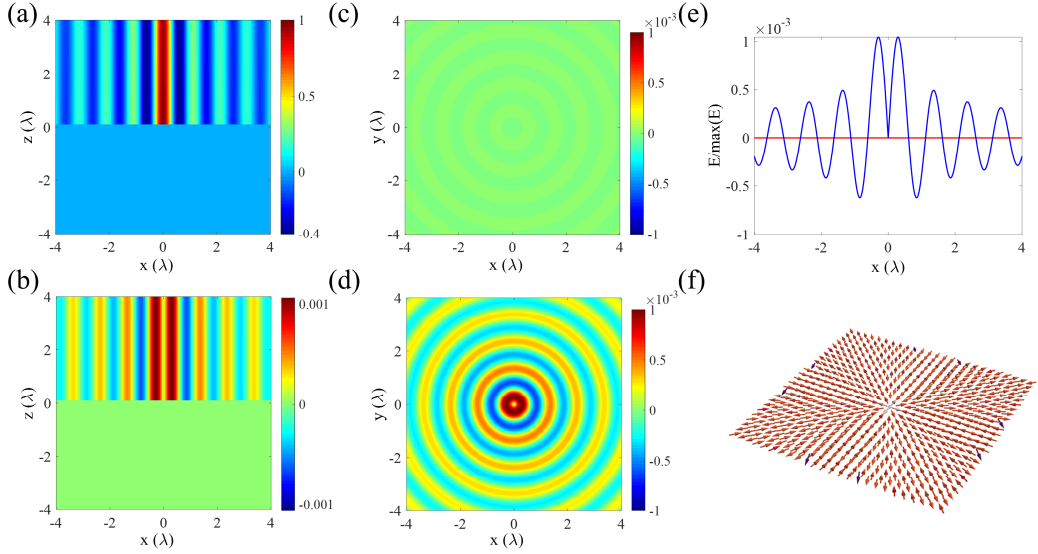

**Fig. S3.** Electric field defect constructed by the 0-order Bessel surface mode at a low frequency limit. The real parts of (a) *z*- and (b) radial electric field components in *xz*-plane (*y* = 0). In the metal layer, the real parts of (c) *z*- and (d) radial electric field components in *xy*-plane (*z* = *a*/2 − 10nm), and the corresponding (e) 1D contour at *y* = 0 and (f) electric field texture (region: 8λ×8λ). One can find that the amplitude of radial component electric field is much larger that of *z*-component electric field, which makes that the skyrmion number of electric field texture is approximatively zero ($N_{SK}$ = 0). Here, the wavelength λ = 1×10$^{-3}$m. The angular frequency ω = 1.88×10$^{12}$/s. *a* is the thickness of metal and here we set *a* = 0.2λ.

In the following, we analyze the electric field topological defects in the multilayered air-metal-air structure. We primarily take the 0-order Bessel-type surface mode, which was known as electric field skyrmions in various references [18,20], for instance. We must emphasize that the results can be generalized into other electric field topological defects because these topological defects can be constructed with the superposition of Bessel-type mode arranged in a specific symmetry [26].

If the frequency is small enough (such as terahertz wave with ω << $ω_p$), one can find that 1 + $ε^m/ε_0$ ≈ $ε^m/ε_0$ and the propagating constant β ≈ $k^±$ = $k_0$ is a pure real number approximatively. In the case, the decaying factors in the upper/lower subspace are $k_z^± = \sqrt{β^2 - (k^±)^2} ≈ 0$. The horizontal electric field components are zero approximatively (**Fig. S3(a-b)**), especially at a low frequency limit, and thus the topological defects constructed by the electric field of 0-order Bessel-type surface mode are considered as a combination of purely 'up' state and 'down' state. While in the metal layer, there is $k_z^m = \sqrt{β^2 - (k^m)^2} \gg β$. The horizontal electric field components are much larger than the normal electric field component (**Fig. S3(c-d)**). At a low frequency limit, the normal electric field component can be ignored and thus the skyrmion number of the topological defects constructed by the electric field of 0-order Bessel-type surface mode can be considered as zero ($N_{SK}$ = 0) (**Fig. S3(e-f)**). In addition, the electric field skyrmion lattice constructed by this SPP mode in C6 symmetry can be found in **Fig. 3(a)** in main text, whose skyrmion number is also considered as zero ($N_{SK}$ = 0).

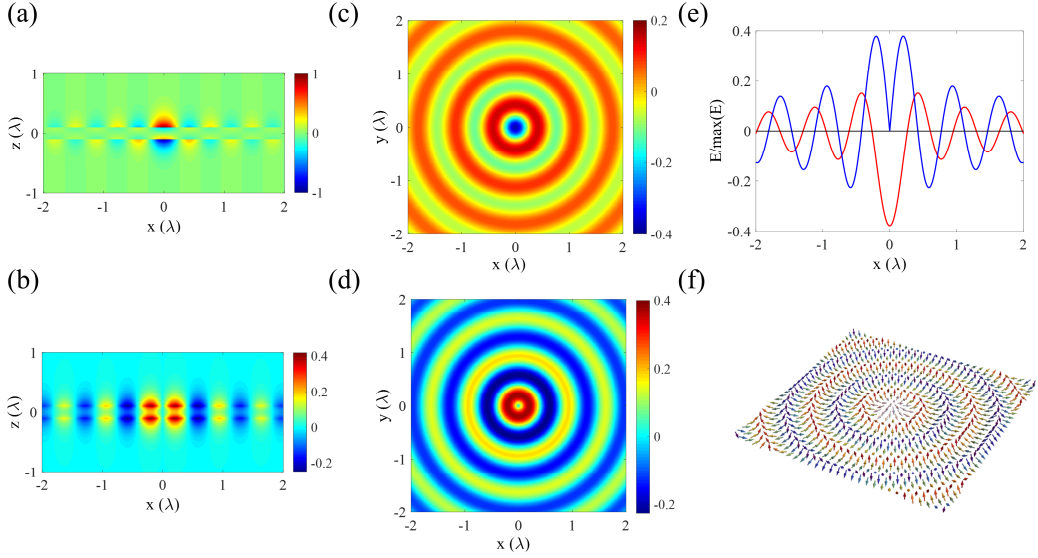

**Fig. S4.** Electric field defect constructed by the 0-order Bessel-type surface mode at visible frequency. The real parts of (a) *z*- and (b) radial electric field components in *xz*-plane (*y* = 0). In the metal layer, the real parts of (c) *z*- and (d) radial electric field components in *xy*-plane (*z* = *a*/2 − 10nm), and the corresponding (e) 1D contour at *y* = 0 and (f) electric field texture (region: 3λ×3λ). For the electric field texture in the upper/lower subspace or in the metal layer, one can also find that the electric field vector whirls from 'down' state to 'up' state along the radial direction, which is a manifestation Néel-type topological texture. Here, $\lambda = 6.38\times10^{-7}m$, which is corresponding to the angular frequency $\omega = 2.95\times10^{15}/s$. *a* is the thickness of metal and here we set $a = 0.2\lambda$.

As the frequency increases (for example, in the near-infrared or visible region), the propagating constant is a complex number and $\text{Re}(\beta) > k_0$ owing to the fact that $\text{Re}(\varepsilon^m/\varepsilon_0) < \text{Re}(1+\varepsilon^m/\varepsilon_0) < 0$. In the instance, the *z*-component wavenumber $k_z^m = \sqrt{\beta^2 - (k^m)^2}$ has a same order of magnitudes with the propagating constant $\beta$. The azimuthal electric field component is always zero for the *p*-polarized 0-order Bessel-type surface mode and the electric field vector varies from the 'down' state to 'up' state in the radial direction (Néel-type) (**Fig. S4**). Therefore, the electric field vector of field skyrmion lattice constructed by this lossy SPP mode in C6 symmetry (**Fig. 3(b)** in main text) whirls along the radial direction and its skyrmion number is +1 ($N_{SK}$ = +1) in the metal layer. Noteworthily, owing to the boundary condition from Maxwell's equations, the normal electric field components are inverted and the horizontal electric field components are continuous through the interface.

At the lossless limit, the topological properties of electric field topological defects are similar to those of lossy metal (**Fig. S5**), which indicates that the lossy property of material does not affect the topological geometries of these electric field topological defects. Thus, the skyrmion number of the electric field skyrmion lattice constructed by this lossless SPP mode in C6 symmetry is also +1 (**Fig. 3(b)** in main text).

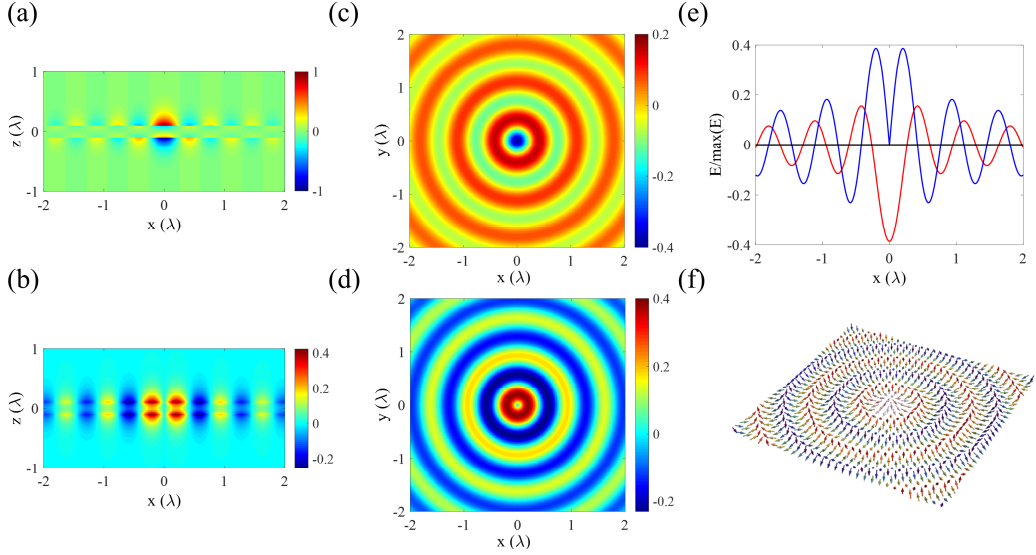

**Fig. S5.** Electric field defect constructed by the 0-order Bessel-type surface mode at the lossless limit. The real parts of (a) *z*- and (b) radial electric field components in *xz*-plane (*y* = 0). In metal layer, the real parts of (c) *z*- and (d) radial electric field components in *xy*-plane (*z* = *a*/2 − 10nm), and the corresponding (e) 1D contour at *y* = 0 and (f) electric field texture (region: 3λ×3λ). One can also find that the electric field vector whirls from 'down' state to 'up' state along the radial direction, which is a manifestation Néel-type topological texture. Here, $\gamma = 0$, $\lambda = 6.38 \times 10^{-7}$m, which is corresponding to the angular frequency $\omega = 2.95 \times 10^{15}$/s. *a* is the thickness of metal and here we set $a = 0.2\lambda$.

Furthermore, as the frequency increases further and approaches the plasma frequency $\omega_p$ ($\omega_p/\sqrt{2} < \omega < \omega_p$), the imaginary part of permittivity can be ignored and there will be $1 + \varepsilon^m/\varepsilon_0 > 0$ and $\varepsilon^m/\varepsilon_0 < 0$. In the case, the propagating constant $\beta$ is a pure imaginary number and the propagating mode is forbidden, which was known as band gap as indicated by part **IV** in **Fig. 1(a)**.

## 2. Electric field topological quasiparticles in slot waveguide modes

If the frequency is larger than the plasma frequency $\omega > \omega_p$, the loss can be ignored and the permittivity is

$$0 < \varepsilon(\omega) = \varepsilon_0 \left(1 - \frac{\omega_p^2}{\omega^2}\right) < \varepsilon_0. \tag{S59}$$

The relative permittivity can be found in **Fig. S6(a)**.

In the case, the slot-waveguide-like modes (contain symmetric TE and TM modes) will be excited in the aire-medium-air structure. We must emphasize that, to produce the slot-waveguide-like modes, there should be another two interfaces as shown in Ref. [62]. Here, we assume the two interfaces localized at infinity. Although this will make the total energy density in air infinite, it would not affect the field distributions in the layer or near the air/medium interfaces. The electric/magnetic field components of the slot-waveguide-like modes in Cartesian coordinates (*x*, *y*, *z*) are summarized in **Table S2**.

**Table S2.** Electric/Magnetic field components of the **slot-waveguide-like** modes

|  | $z > +\frac{a}{2}$ | $-\frac{a}{2} < z < +\frac{a}{2}$ | $z < -\frac{a}{2}$ |
| --- | --- | --- | --- |
| $E_x$ | $+C^{E+}\left(-\dfrac{W^E}{\beta^{E2}}\right)\dfrac{\partial \xi^E}{\partial x}\sin(W^E z)$ $+C^{H+}\dfrac{i\omega\mu^+}{\beta^{H2}}\dfrac{\partial \xi^H}{\partial y}\cos(W^H z)$ | $+A^{Em}\left(+\dfrac{U^E}{\beta^{E2}}\right)\dfrac{\partial \xi^E}{\partial x}e^{+U^E z}+A^{Hm}\dfrac{i\omega\mu^m}{\beta^{H2}}\dfrac{\partial \xi^H}{\partial y}e^{+U^H z}$ $+B^{Em}\left(-\dfrac{U^E}{\beta^{E2}}\right)\dfrac{\partial \xi^E}{\partial x}e^{-U^E z}+B^{Hm}\dfrac{i\omega\mu^m}{\beta^{H2}}\dfrac{\partial \xi^H}{\partial y}e^{-U^H z}$ | $+D^{E-}\left(-\dfrac{W^E}{\beta^{E2}}\right)\dfrac{\partial \xi^E}{\partial r}\sin(W^E z)$ $+D^{H-}\dfrac{i\omega\mu^-}{\beta^{H2}}\dfrac{\partial \xi^H}{\partial y}\cos(W^H z)$ |
| $E_y$ | $+C^{E+}\left(-\dfrac{W^E}{\beta^{E2}}\right)\dfrac{\partial \xi^E}{\partial y}\sin(W^E z)$ $-C^{H+}\dfrac{i\omega\mu^+}{\beta^{H2}}\dfrac{\partial \xi^H}{\partial x}\cos(W^H z)$ | $+A^{Em}\left(+\dfrac{U^E}{\beta^{E2}}\right)\dfrac{\partial \xi^E}{\partial y}e^{+U^E z}-A^{Hm}\dfrac{i\omega\mu^m}{\beta^{H2}}\dfrac{\partial \xi^H}{\partial x}e^{+U^H z}$ $+B^{Em}\left(-\dfrac{U^E}{\beta^{E2}}\right)\dfrac{\partial \xi^E}{\partial y}e^{-U^E z}-B^{Hm}\dfrac{i\omega\mu^m}{\beta^{H2}}\dfrac{\partial \xi^H}{\partial x}e^{-U^H z}$ | $+D^{E-}\left(-\dfrac{W^E}{\beta^{E2}}\right)\dfrac{\partial \xi^E}{\partial y}\sin(W^E z)$ $-D^{H-}\dfrac{i\omega\mu^-}{\beta^{H2}}\dfrac{\partial \xi^H}{\partial x}\cos(W^H z)$ |
| $E_z$ | $C^{E+}\xi^E\cos(W^E z)$ | $A^{Em}\xi^E e^{+U^E z}+B^{Em}\xi^E e^{-U^E z}$ | $D^{E-}\xi^E\cos(W^E z)$ |
| $H_x$ | $-C^{E+}\dfrac{i\omega\varepsilon^+}{\beta^{E2}}\dfrac{\partial \xi^E}{\partial y}\cos(W^E z)$ $+C^{H+}\left(-\dfrac{W^H}{\beta^{H2}}\right)\dfrac{\partial \xi^H}{\partial x}\sin(W^H z)$ | $-A^{Em}\dfrac{i\omega\varepsilon^m}{\beta^{E2}}\dfrac{\partial \xi^E}{\partial y}e^{+U^E z}+A^{Hm}\left(+\dfrac{U^H}{\beta^{H2}}\right)\dfrac{\partial \xi^H}{\partial x}e^{+U^H z}$ $-B^{Em}\dfrac{i\omega\varepsilon^m}{\beta^{E2}}\dfrac{\partial \xi^E}{\partial y}e^{-U^E z}+B^{Hm}\left(-\dfrac{U^H}{\beta^{H2}}\right)\dfrac{\partial \xi^H}{\partial x}e^{-U^H z}$ | $-D^{E-}\dfrac{i\omega\varepsilon^-}{\beta^{E2}}\dfrac{\partial \xi^E}{\partial y}\cos(W^E z)$ $+D^{H-}\left(-\dfrac{W^H}{\beta^{H2}}\right)\dfrac{\partial \xi^H}{\partial x}\sin(W^H z)$ |
| $H_y$ | $+C^{E+}\dfrac{i\omega\varepsilon^+}{\beta^{E2}}\dfrac{\partial \xi^E}{\partial x}\cos(W^E z)$ $+C^{H+}\left(-\dfrac{W^H}{\beta^{H2}}\right)\dfrac{\partial \xi^H}{\partial y}\sin(W^H z)$ | $+A^{Em}\dfrac{i\omega\varepsilon^m}{\beta^{E2}}\dfrac{\partial \xi^E}{\partial x}e^{+U^E z}+A^{Hm}\left(+\dfrac{U^H}{\beta^{H2}}\right)\dfrac{\partial \xi^H}{\partial y}e^{+U^H z}$ $+B^{Em}\dfrac{i\omega\varepsilon^m}{\beta^{E2}}\dfrac{\partial \xi^E}{\partial x}e^{-U^E z}+B^{Hm}\left(-\dfrac{U^H}{\beta^{H2}}\right)\dfrac{\partial \xi^H}{\partial y}e^{-U^H z}$ | $+D^{E-}\dfrac{i\omega\varepsilon^-}{\beta^{E2}}\dfrac{\partial \xi^E}{\partial x}\cos(W^E z)$ $+D^{H-}\left(-\dfrac{W^H}{\beta^{H2}}\right)\dfrac{\partial \xi^H}{\partial y}\sin(W^H z)$ |
| $H_z$ | $C^{H+}\xi^H\cos(W^H z)$ | $A^{Hm}\xi^H e^{+U^H z}+B^{Hm}\xi^H e^{-U^H z}$ | $D^{H-}\xi^H\cos(W^H z)$ |

Therein, the electric and magnetic Hertz potentials fulfil the transverse Helmholtz equations:

$$\nabla_\perp^2 \xi^E + \beta^{E2}\xi^E = 0 \quad \nabla_\perp^2 \xi^H + \beta^{H2}\xi^H = 0$$

in the layer and the surround materials. The superscript $E$ and $H$ represent the electric modes (TM modes) and magnetic (transverse electric: TE modes) modes, respectively. By considering the boundary conditions, the dispersion relation can be expressed as

$$\frac{\varepsilon^m}{U^E}\coth\left(\frac{U^E a}{2}\right) = -\frac{\varepsilon^\pm}{W^E}\cot\left(\frac{W^E a}{2}\right) \tag{S60}$$

for the electric mode, and

$$\frac{\mu^m}{U^H}\coth\left(\frac{U^H a}{2}\right) = -\frac{\mu^\pm}{W^H}\cot\left(\frac{W^H a}{2}\right) \tag{S61}$$

for the magnetic mode. Here, the parameters $U^E$, $W^E$, $U^H$ and $W^H$ can be calculated by

$$\begin{aligned} W^{E2} = k^{\pm 2} - \beta^{E2} & \quad W^{H2} = k^{\pm 2} - \beta^{H2} \\ U^{E2} = \beta^{E2} - k^{m2} & \quad U^{H2} = \beta^{H2} - k^{m2} \end{aligned}. \tag{S62}$$

By solving the propagation constants $\beta^E$ and $\beta^H$ (pure real numbers), the field parameters are

$$\begin{aligned} A^{Em} = B^{Em} \quad C^{E+} = D^{E-} \quad A^{Em} = C^{E+}\frac{\varepsilon^+}{\varepsilon^m}\cos\left(+\frac{W^E a}{2}\right)\Big/\left[\exp\left(+\frac{U^E a}{2}\right)+\exp\left(-\frac{U^E a}{2}\right)\right] \\ A^{Hm} = B^{Hm} \quad C^{H+} = D^{H-} \quad A^{Hm} = C^{H+}\frac{\mu^+}{\mu^m}\cos\left(+\frac{W^H a}{2}\right)\Big/\left[\exp\left(+\frac{U^H a}{2}\right)+\exp\left(-\frac{U^H a}{2}\right)\right] \end{aligned}. \tag{S63}$$

Since only the relative amplitude makes physical sense, one can set $A^{Hm}$ to be 1 and $A^{Em} = \pm i\sqrt{\mu_0/\varepsilon_0}$, and then the field distributions excited by circularly polarized light can be calculated properly. The dispersion relations of the symmetric TM and TE modes are shown in the red and blue lines of **Fig. S6(b-c)**, respectively.

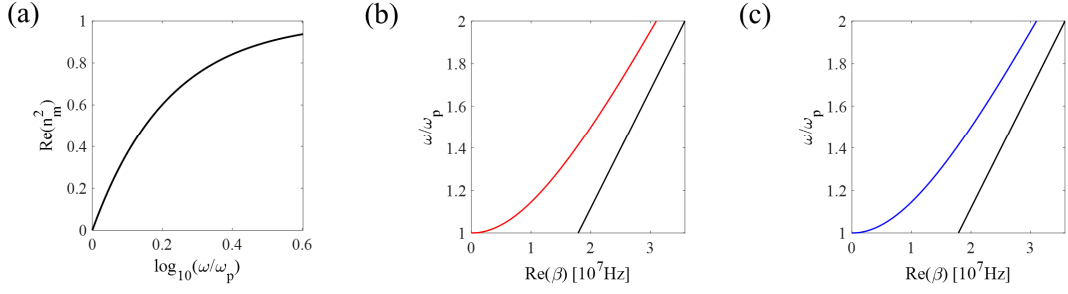

**Fig. S6.** Relative permittivity and dispersive relations of slot-waveguide-like modes at air-medium-air structure. (a) the relative permittivity of medium via the frequency in the situation $\omega > \omega_p$; the propagating constants of (b) symmetric electric modes and (c) symmetric magnetic modes via the frequency. Since the slot-waveguide-like mode is occupied in the whole space, the propagating constants of these symmetric electric and magnetic modes are equal. The black lines represent the light cones in air. The thickness of layer is equal to $0.2\lambda$. $\omega_p \approx 5.36\times 10^{15}/s$ ($\lambda_p \approx 0.35\times 10^{-6} m$) is the plasma frequency and the characteristic frequency $\gamma = 10^{14}/s$.

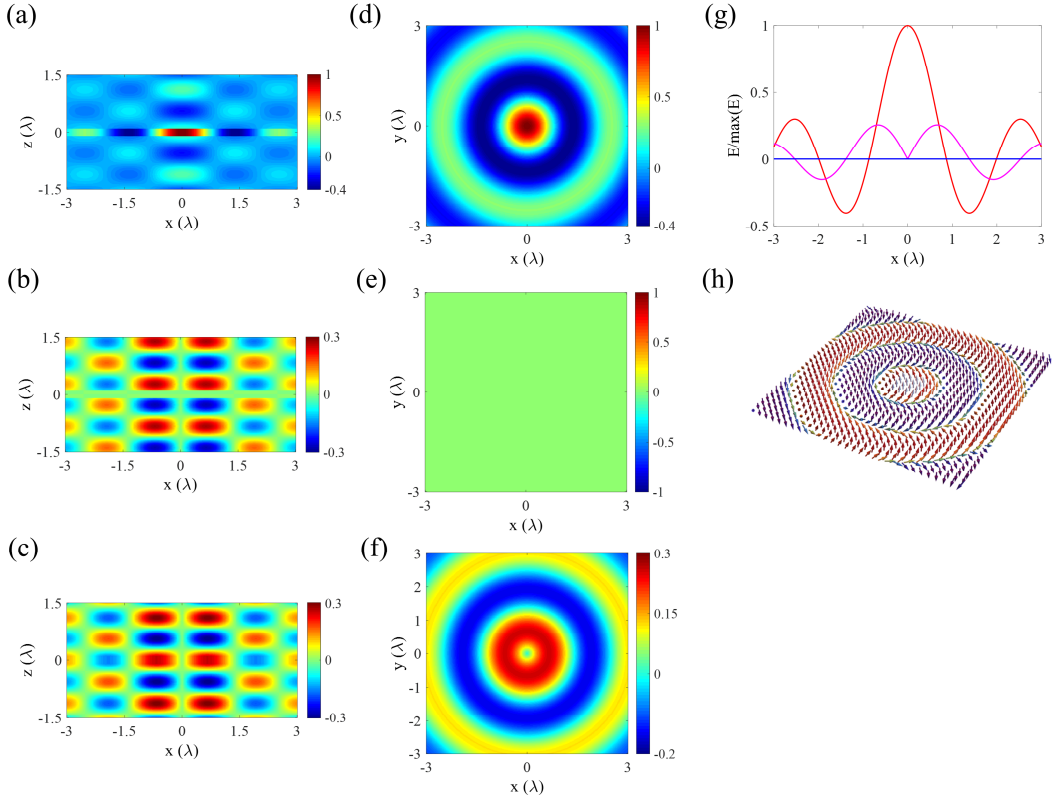

**Fig. S7.** Electric field defect constructed by the 0-order Bessel-type surface mode for $\omega > \omega_p$. The imaginary parts of (a) $z$-, (b) radial and (c) azimuthal electric field components in $xz$-plane ($y = 0$). In the layer, the imaginary parts of (d) $z$-, (e) radial and (f) azimuthal electric field components in $xy$-plane ($z = 0$), and the corresponding (g) 1D contour at $y = 0$ and (m) electric field texture (region: $8\lambda \times 8\lambda$). It is worth noting that one can transform the imaginary part into the real part by add an additional constant phase to the amplitude. It can be found that, in the medium layer, the electric field vector whirls from 'up' state to 'down' state along the azimuthal direction, which is a manifestation of Bloch-type topological texture. Here, $\lambda = 3.33\times 10^{-7}m$, which is corresponding to the angular frequency $\omega = 5.66\times 10^{15}/s$. $a$ is the thickness of metal and here we set $a = 0.2\lambda$.

In the case, since the slot-waveguide-like mode is occupied in the whole space, the propagating constants of symmetric electric and magnetic modes are equal approximatively. In air, the azimuthal and radial electric field components exist simultaneously for the 0-order Bessel-type surface mode (**Fig. S7(a-c)**). The electric field vector varies from the 'up/down' state to 'down/up' state along the radial and azimuthal directions simultaneously, and thus the electric field textures can be considered as twisted configurations in the upper/lower subspace. While in the medium layer, the radial electric field component cancels out due to the mode's coupling (**Fig. S7(b,e)**). Thus, the electric field topological defect in the layer can be considered as Bloch-type configuration (**Fig. S7(g-h)**). And the skyrmion number of this electric field skyrmion lattice constructed by this slot-waveguide-like mode in C6 symmetry is −1 (**Fig. 3(c)**). Noteworthily, owing to the boundary condition from Maxwell's equations, the normal electric field components are parallel and the horizontal electric field components are continuous through the interface (**Fig. S7**).

## 3. Electric field topological quasiparticles in dielectric waveguide wave mode

For the dielectrics, the damping can be ignored naturally the permittivity is (**Fig. S8(a)**)

$$\varepsilon(\omega) = \varepsilon_0 \left( 1 + \sum_i \frac{f_i \omega_p^2}{\omega_i^2 - \omega^2} \right). \tag{S64}$$

The relative permittivity can be found in **Fig. S8(a)**.

In the case, the waveguide modes (contain symmetric and anti-symmetric TE and TM modes) will be produced in the dielectric layer. The electric/magnetic field components of the waveguide modes in Cartesian coordinates ($x, y, z$) are summarized in **Table S3**.

**Table S3.** Electric/Magnetic field components of the **waveguide** modes

| | $z > +\frac{a}{2}$ | $-\frac{a}{2} < z < +\frac{a}{2}$ | $z < -\frac{a}{2}$ |
|---|---|---|---|
| $E_x$ | $+C^{E+}\left(-\frac{W^E}{\beta^{E2}}\right)\frac{\partial \xi^E}{\partial x}e^{-W^E z}$ $+C^{H+}\frac{i\omega\mu^+}{\beta^{H2}}\frac{\partial \xi^H}{\partial y}e^{-W^H z}$ | $+A^{Em}\left(+\frac{iU^E}{\beta^{E2}}\right)\frac{\partial \xi^E}{\partial x}e^{+iU^E z} + A^{Hm}\frac{i\omega\mu^m}{\beta^{H2}}\frac{\partial \xi^H}{\partial y}e^{+iU^H z}$ $+B^{Em}\left(-\frac{iU^E}{\beta^{E2}}\right)\frac{\partial \xi^E}{\partial x}e^{-iU^E z} + B^{Hm}\frac{i\omega\mu^m}{\beta^{H2}}\frac{\partial \xi^H}{\partial y}e^{-iU^H z}$ | $+D^{E-}\left(-\frac{W^E}{\beta^{E2}}\right)\frac{\partial \xi^E}{\partial r}e^{+W^E z}$ $+D^{H-}\frac{i\omega\mu^-}{\beta^{H2}}\frac{\partial \xi^H}{\partial y}e^{+W^H z}$ |
| $E_y$ | $+C^{E+}\left(-\frac{W^E}{\beta^{E2}}\right)\frac{\partial \xi^E}{\partial y}e^{-W^E z}$ $-C^{H+}\frac{i\omega\mu^+}{\beta^{H2}}\frac{\partial \xi^H}{\partial x}e^{-W^H z}$ | $+A^{Em}\left(+\frac{iU^E}{\beta^{E2}}\right)\frac{\partial \xi^E}{\partial y}e^{+iU^E z} - A^{Hm}\frac{i\omega\mu^m}{\beta^{H2}}\frac{\partial \xi^H}{\partial x}e^{+iU^H z}$ $+B^{Em}\left(-\frac{iU^E}{\beta^{E2}}\right)\frac{\partial \xi^E}{\partial y}e^{-iU^E z} - B^{Hm}\frac{i\omega\mu^m}{\beta^{H2}}\frac{\partial \xi^H}{\partial x}e^{-iU^H z}$ | $+D^{E-}\left(-\frac{W^E}{\beta^{E2}}\right)\frac{\partial \xi^E}{\partial y}e^{+W^E z}$ $-D^{H-}\frac{i\omega\mu^-}{\beta^{H2}}\frac{\partial \xi^H}{\partial x}e^{+W^H z}$ |
| $E_z$ | $C^{E+}\xi^E e^{-W^E z}$ | $A^{Em}\xi^E e^{+iU^E z} + B^{Em}\xi^E e^{-iU^E z}$ | $D^{E-}\xi^E e^{+W^E z}$ |
| $H_x$ | $-C^{E+}\frac{i\omega\varepsilon^+}{\beta^{E2}}\frac{\partial \xi^E}{\partial y}e^{-W^E z}$ $+C^{H+}\left(-\frac{W^H}{\beta^{H2}}\right)\frac{\partial \xi^H}{\partial x}e^{-W^H z}$ | $-A^{Em}\frac{i\omega\varepsilon^m}{\beta^{E2}}\frac{\partial \xi^E}{\partial y}e^{+iU^E z} + A^{Hm}\left(+\frac{iU^H}{\beta^{H2}}\right)\frac{\partial \xi^H}{\partial x}e^{+iU^H z}$ $-B^{Em}\frac{i\omega\varepsilon^m}{\beta^{E2}}\frac{\partial \xi^E}{\partial y}e^{-iU^E z} + B^{Hm}\left(-\frac{iU^H}{\beta^{H2}}\right)\frac{\partial \xi^H}{\partial x}e^{-iU^H z}$ | $-D^{E-}\frac{i\omega\varepsilon^-}{\beta^{E2}}\frac{\partial \xi^E}{\partial y}e^{+W^E z}$ $+D^{H-}\left(-\frac{W^H}{\beta^{H2}}\right)\frac{\partial \xi^H}{\partial x}e^{+W^H z}$ |
| $H_y$ | $+C^{E+}\frac{i\omega\varepsilon^+}{\beta^{E2}}\frac{\partial \xi^E}{\partial x}e^{-W^E z}$ $+C^{H+}\left(-\frac{W^H}{\beta^{H2}}\right)\frac{\partial \xi^H}{\partial y}e^{-W^H z}$ | $+A^{Em}\frac{i\omega\varepsilon^m}{\beta^{E2}}\frac{\partial \xi^E}{\partial x}e^{+iU^E z} + A^{Hm}\left(+\frac{iU^H}{\beta^{H2}}\right)\frac{\partial \xi^H}{\partial y}e^{+iU^H z}$ $+B^{Em}\frac{i\omega\varepsilon^m}{\beta^{E2}}\frac{\partial \xi^E}{\partial x}e^{-iU^E z} + B^{Hm}\left(-\frac{iU^H}{\beta^{H2}}\right)\frac{\partial \xi^H}{\partial y}e^{-iU^H z}$ | $+D^{E-}\frac{i\omega\varepsilon^-}{\beta^{E2}}\frac{\partial \xi^E}{\partial x}e^{+W^E z}$ $+D^{H-}\left(-\frac{W^H}{\beta^{H2}}\right)\frac{\partial \xi^H}{\partial y}e^{+W^H z}$ |
| $H_z$ | $C^{H+}\xi^H e^{-W^H z}$ | $A^{Hm}\xi^H e^{+iU^H z} + B^{Hm}\xi^H e^{-iU^H z}$ | $D^{H-}\xi^H e^{+W^H z}$ |

Therein, the electric and magnetic Hertz potentials fulfil the transverse Helmholtz equations:

$$\nabla_\perp^2 \xi^E + \beta^{E2} \xi^E = 0 \quad \nabla_\perp^2 \xi^H + \beta^{H2} \xi^H = 0 \tag{S65}$$

in the metal region and surround materials. The superscripts $E$ and $H$ represent the electric modes and magnetic modes, respectively. By considering the boundary conditions, the dispersion relations can be expressed as

$$\frac{\varepsilon^m}{U^E}\cot\left(\frac{U^E a}{2}\right) = +\frac{\varepsilon^{\pm}}{W^E} \tag{S66}$$

for the anti-symmetric electric mode, and

$$\frac{\varepsilon^m}{U^E}\tan\left(\frac{U^E a}{2}\right) = -\frac{\varepsilon^{\pm}}{W^E} \tag{S67}$$

for the symmetric electric mode, and

$$\frac{\mu^m}{U^H}\cot\left(\frac{U^H a}{2}\right) = +\frac{\mu^{\pm}}{W^H} \tag{S68}$$

for the anti-symmetric magnetic mode, and

$$\frac{\mu^m}{U^H}\tan\left(\frac{U^H a}{2}\right) = -\frac{\mu^{\pm}}{W^H} \tag{S69}$$

for the symmetric magnetic mode. Here, the parameters $U^E$, $W^E$, $U^H$ and $W^H$ can be expressed as

$$\begin{array}{ll} W^{E2} = \beta^{E2} - k^{\pm 2} & W^{H2} = \beta^{H2} - k^{\pm 2} \\ U^{E2} = k^{m2} - \beta^{E2} & U^{H2} = k^{m2} - \beta^{H2} \end{array}. \tag{S70}$$

By solving the propagation constants $\beta^E$ and $\beta^H$ (pure real numbers), the field parameters for the symmetric modes are

$$\begin{array}{lll} A^{Em} = B^{Em} & C^{E+} = D^{E-} & A^{Em} = C^{E+}\dfrac{\varepsilon^{\pm}}{\varepsilon^m}\exp\left(-\dfrac{W^E a}{2}\right) \Big/ \left[\exp\left(+i\dfrac{U^E a}{2}\right) + \exp\left(-i\dfrac{U^E a}{2}\right)\right] \\[2mm] A^{Hm} = B^{Hm} & C^{H+} = D^{H-} & A^{Hm} = C^{H+}\dfrac{\mu^{\pm}}{\mu^m}\exp\left(-\dfrac{W^H a}{2}\right) \Big/ \left[\exp\left(+\dfrac{iU^H a}{2}\right) + \exp\left(-\dfrac{iU^H a}{2}\right)\right] \end{array}. \tag{S71}$$

Moreover, the field parameters for the anti-symmetric modes are

$$\begin{array}{lll} A^{Em} = +B^{Em} & C^{E+} = -D^{E-} & A^{Em} = C^{E+}\dfrac{\varepsilon^{\pm}}{\varepsilon^m}\exp\left(-\dfrac{W^E a}{2}\right) \Big/ \left[\exp\left(+i\dfrac{U^E a}{2}\right) - \exp\left(-i\dfrac{U^E a}{2}\right)\right] \\[2mm] A^{Hm} = +B^{Hm} & C^{H+} = -D^{H-} & A^{Hm} = C^{H+}\dfrac{\mu^{\pm}}{\mu^m}\exp\left(-\dfrac{W^H a}{2}\right) \Big/ \left[\exp\left(+\dfrac{iU^H a}{2}\right) - \exp\left(-\dfrac{iU^H a}{2}\right)\right] \end{array}. \tag{S72}$$

Since only the relative amplitude makes physical sense, one can set $A^{Hm}$ to be 1 and $A^{Em} = \pm i\sqrt{\mu_0/\varepsilon_0}$, and then the field distributions excited by circularly polarized light can be calculated properly. The dispersion relations of the symmetric (red/magenta) and antisymmetric (blue/green) magnetic/electric modes are shown in **Fig. S8(b-c)**, respectively.

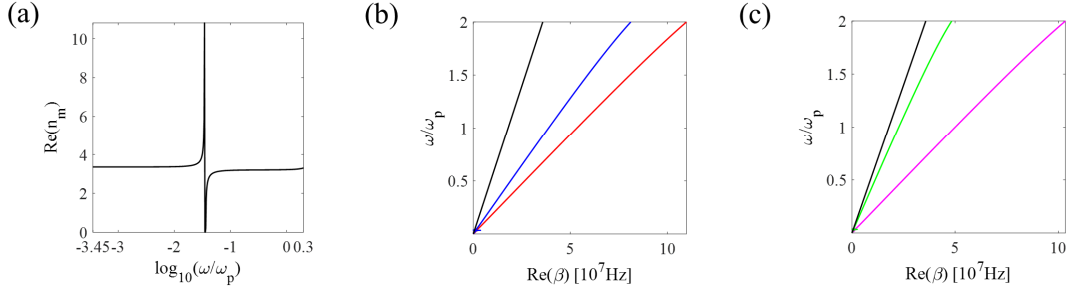

**Fig. S8.** Dispersive relations of waveguide modes at air-dielectric-air structure. (a) the relative permittivity of dielectric via the frequency; (b) the propagating constants of symmetric (red line) and anti-symmetric (blue line) magnetic modes via the frequency and (c) the propagating constants of symmetric (magenta line) and anti-symmetric (green line) electric modes via the frequency. Here, we only focus on the frequency with refractive index larger than 1. And the relative permittivity is $n^2 = 9 + \frac{\frac{A_1}{B_1}4\pi^2c^2}{\frac{1}{B_1}4\pi^2c^2-\omega^2} + \frac{\frac{A_2}{B_2}4\pi^2c^2}{\frac{1}{B_2}4\pi^2c^2-\omega^2} + \frac{\frac{A_3}{B_3}4\pi^2c^2}{\frac{1}{B_3}4\pi^2c^2-\omega^2}$, where $A_1$=1.03961212, $B_1$=0.00600069867, $A_2$=0.231792344, $B_2$=0.0200179144, $A_3$=1.01046945 and $B_3$=103.560653.

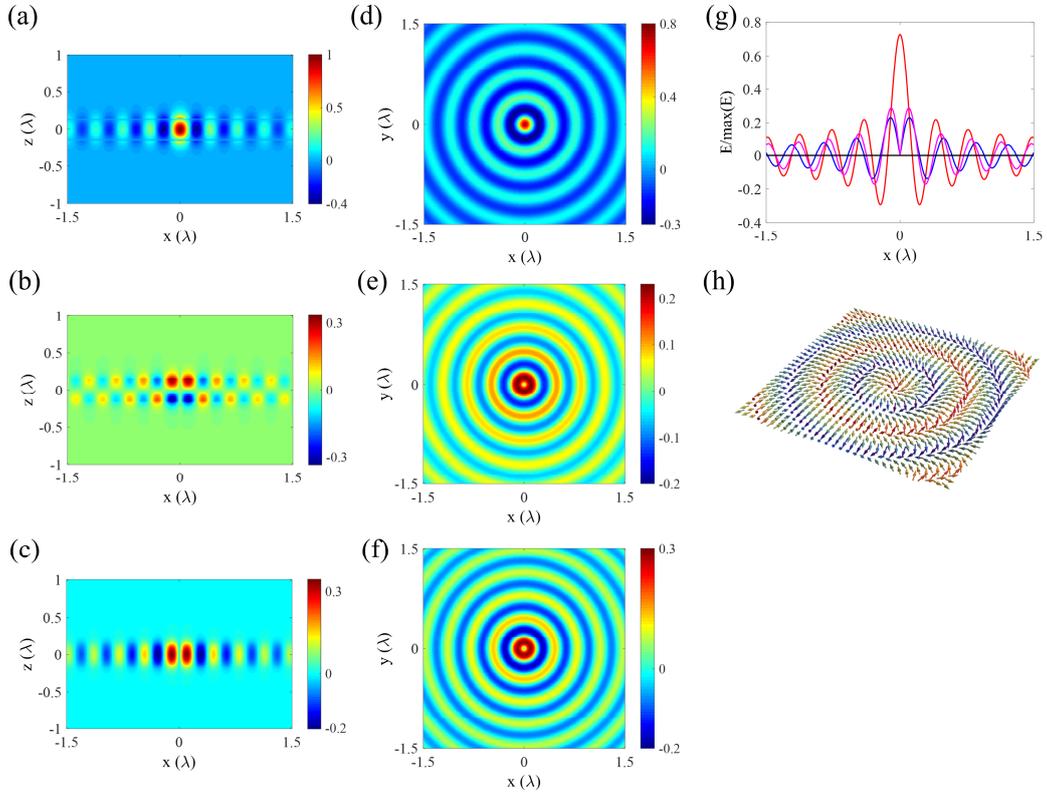

**Fig. S9.** Electric field defect constructed by the 0-order Bessel-type guided wave mode. The imaginary parts of (a) $z$-, (b) radial and (c) azimuthal electric field components in $xz$-plane ($y = 0$). In the layer, the imaginary parts of of (d) $z$-, (e) radial and (f) azimuthal electric field components in $xy$-plane ($z = a/2 - 10$nm), and the corresponding (g) 1D contour at $y = 0$ and (h) electric field texture (region: $1.2\lambda \times 1.2\lambda$). One can find that the electric field vector whirls from 'up' state to 'down' state along the radial and azimuthal direction simultaneously, which is a manifestation of twisted-type topological texture. Here, $\lambda = 9.90 \times 10^{-7}$m, which is corresponding to the angular frequency $\omega = 1.90 \times 10^{15}$/s. $a$ is the thickness of metal and here we set $a = 0.3\lambda$.

In the case, the propagating constants of symmetric and anti-symmetric electric/magnetic modes are different completely. In the layer, the radial and azimuthal electric field components exist simultaneously for the 0-order Bessel-type mode, which is a manifestation of twisted-type field textures (**Fig. S9(d-h)**). The skyrmion number of electric field skyrmion lattice constructed by this waveguide mode can be −1, as shown in **Fig. 3(d)** in the main text. Noteworthily, owing to the boundary condition from Maxwell's equations, the normal electric field components are parallel and the horizontal electric field components are continuous through the interface (**Fig. S9**).

## V. EM spin topological defects and their topological state transitions

In the section VI, we analyze the mode properties and electric field topological textures in multilayered structure. The formations of various configurations of field topological textures are based on the modes' coupling but not the SOCs. On the other hand, to understand the formations of various photonic spin topological defects in multilayered structures, we should analyze the SOC terms given in the Section II. Noteworthily, the dispersive effect is not considered in SOC terms (Eq. (S29) and Eq. (S31)) [28,56].

In the layer, the real and imaginary parts of SOC term are given by the gradient of helical density and vorticites of kinetic Abraham-Poynting momentum. The expressions of gradient helical density are

$$\left(\text{Re}\,\mathbf{H}_{\text{SO}}\right)_r = -\frac{\hbar\omega}{k^2}(\nabla C)_r = \frac{\hbar}{2}\text{Im}\left\{\frac{\partial E_z^*}{\partial r}H_z + E_z^*\frac{\partial H_z}{\partial r} + \frac{\partial E_r^*}{\partial r}H_r + E_r^*\frac{\partial H_r}{\partial r} + \frac{\partial E_\varphi^*}{\partial r}H_\varphi + E_\varphi^*\frac{\partial H_\varphi}{\partial r}\right\}, \tag{S73}$$

$$\left(\text{Re}\,\mathbf{H}_{\text{SO}}\right)_\varphi = -\frac{\hbar\omega}{k^2}(\nabla C)_\varphi = 0, \tag{S74}$$

and

$$\left(\text{Re}\,\mathbf{H}_{\text{SO}}\right)_z = -\frac{\hbar\omega}{k^2}(\nabla C)_z = \frac{\hbar}{2}\text{Im}\left\{\frac{\partial E_z^*}{\partial z}H_z + E_z^*\frac{\partial H_z}{\partial z} + \frac{\partial E_r^*}{\partial z}H_r + E_r^*\frac{\partial H_r}{\partial z} + \frac{\partial E_\varphi^*}{\partial z}H_\varphi + E_\varphi^*\frac{\partial H_\varphi}{\partial z}\right\}. \tag{S75}$$

And the vorticites of kinetic Abraham-Poynting momentum are

$$\left(\text{Im}\,\mathbf{H}_{\text{SO}}\right)_r = \frac{\hbar}{2}\left(\nabla\times\mathbf{p}^A\right)_r = -\frac{\hbar}{2}\frac{\partial p_\varphi^A}{\partial z} = \frac{\hbar}{4}\text{Re}\left\{-\frac{\partial E_z^*}{\partial z}H_r - E_z^*\frac{\partial H_r}{\partial z} + \frac{\partial E_r^*}{\partial z}H_z + E_r^*\frac{\partial H_z}{\partial z}\right\}, \tag{S76}$$

$$\left(\text{Im}\,\mathbf{H}_{\text{SO}}\right)_\varphi = \frac{\hbar}{2}\left(\nabla\times\mathbf{p}^A\right)_\varphi = \frac{\hbar}{2}\left(\frac{\partial p_r^A}{\partial z} - \frac{\partial p_z^A}{\partial r}\right) = \frac{\hbar}{4}\text{Re}\left\{\begin{array}{l}\frac{\partial E_\varphi^*}{\partial z}H_z + E_\varphi^*\frac{\partial H_z}{\partial z} - \frac{\partial E_z^*}{\partial z}H_\varphi - E_z^*\frac{\partial H_\varphi}{\partial z}\\ -\frac{\partial E_r^*}{\partial r}H_\varphi - E_r^*\frac{\partial H_\varphi}{\partial r} + \frac{\partial E_\varphi^*}{\partial r}H_r + E_\varphi^*\frac{\partial H_r}{\partial r}\end{array}\right\}, \tag{S77}$$

$$\left(\text{Im}\,\mathbf{H}_{\text{SO}}\right)_z = \frac{\hbar}{2}\left(\nabla\times\mathbf{p}^A\right)_z = \frac{\hbar}{2}\frac{1}{r}\frac{\partial rp_\varphi^A}{\partial r} = \frac{\hbar}{4}\text{Re}\left\{\frac{\partial E_z^*}{\partial r}H_r + E_z^*\frac{\partial H_r}{\partial r} - \frac{\partial E_r^*}{\partial r}H_z - E_r^*\frac{\partial H_z}{\partial r} + \frac{E_z^*H_r - E_r^*H_z}{r}\right\}. \tag{S78}$$

For the real part of SOC term, the variation of EM helicity is corresponding to the variation of longitudinal spin, because the longitudinal spin is proportional to the EM helicity by $\mathbf{S}_l = \sum \hbar\sigma_h\hat{\mathbf{k}}$ [57]. The variation of longitudinal spin is widely existing in the phenomena of spin-orbit interactions such as: the spin-based position of light and the spin-based orbital angular momentum of light [43,44]. Whereas the vorticities of kinetic Abraham-Poynting momentum is corresponding to the EM transverse spin [23,57,58], which is related to the structural properties of EM fields. Moreover, the EM transverse spin given by the vorticities of kinetic Abraham-Poynting momentum also represents the intrinsic spin-momentum locking of light [60], and is widely existing in the phenomena of spin-orbit interactions such as: unidirectional guided wave of light [36-38], orbital-to-spin AM conversions [43] and photonic spin topological defects [19,21-28]. Particularly, in section III, we demonstrate that either the Néel-type or the Bloch-type or the twisted-type skyrmions can be excited in the confined EM field, and the point that determines which one would be generated is the SOC terms given by Eqs. (S73-S78).

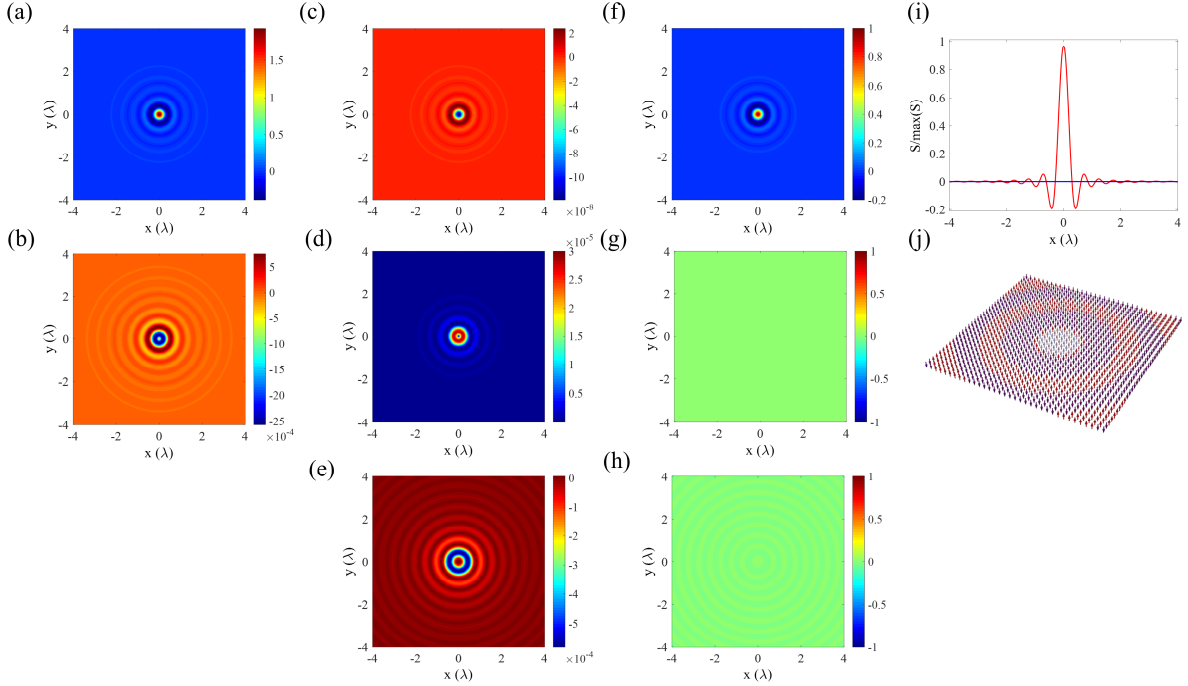

**Fig. S10.** SOC and spin texture of +1-order Bessel-type surface mode at a low frequency limit. In the metal layer, the (a) $z$- and (b) radial components of real part of SOC term Re$\mathbf{H}_{SO}$ in $xy$-plane ($z = a/2 − 10$nm). The (c) $z$-, (d) radial and (e) azimuthal components of imaginary part of SOC term Im$\mathbf{H}_{SO}$ in $xy$-plane ($z = a/2 − 10$nm). The (f) $z$-, (g) radial and (h) azimuthal SAM components in $xy$-plane ($z = a/2 − 10$nm), and the corresponding (i) 1D contour at $y = 0$ and (j) spin texture (region: $2\lambda \times 2\lambda$). One can find that the amplitude of $z$-component SAM is much larger that of horizontal SAM components, and a cylindrical domain wall-like texture (skyrmion number $N_{SK} = −1$) is present. Here, $\lambda = 1\times10^{-3}m$, which is corresponding to the angular frequency $\omega = 1.88\times10^{12}/s$. $a$ is the thickness of metal and here we set $a = 0.2\lambda$.

For the SPP modes in the air-metal-air multilayered structure, if the frequency is small enough ($\omega \ll \omega_p$, such as terahertz wave), there is $1 + \varepsilon^m/\varepsilon_0 \approx \varepsilon^m/\varepsilon_0$ and thus the propagating constant $\beta \approx k^\pm = k_0$ is a pure real number approximatively, as indicated by region **I** in **Fig. 1(a)**. In the case, the decaying factor in the metal is enormous because $k_z^m = \sqrt{\beta^2 - (k^m)^2} \gg \beta$ and the EM field is only localized in the surface of conductor. Therefore, the real part of the $z$-component of SOC term (Re$\mathbf{H}_{SO})_z$ is much larger than the real part of the horizontal components of SOC term (Re$\mathbf{H}_{SO})_r$ (**Fig. S10(a-b)**). In addition, the imaginary parts of SOC term (Im$\mathbf{H}_{SO}$) will affect the horizontal SAMs by two mechanisms: 1. Rashba-like SOC leading to radial SAM by $(\hat{\mathbf{n}} \times P_\varphi^A) \cdot \mathbf{S}$, which is also corresponding to the intrinsic spin-momentum locking of EM wave [60]; 2. Azimuthal SAM originating from the $z$-component kinetic Abraham-Poynting momentum by $(\nabla_r \times P_z^A)$ [58]. The configuration of photonic spin topological defects is determined by the ratio of these two mechanisms. In the instance, the magnitude of $z$-component kinetic Abraham-Poynting momentum is larger than that of azimuthal kinetic Abraham-Poynting momentum, and thus the magnitude of azimuthal EM transverse spin is larger than that of radial EM transverse spin. However, both two the horizontal SAMs are extremely small (**Fig. S10(c-e)**). Thus, it can be considered that there is only the SOC in the z-direction, which results in the z-component SAM being prominent (**Fig. S10(f-h)**). Particularly at a low frequency limit, and the spin topological textures of +1-order Bessel-type surface modes can be considered as a combination of purely 'up' state and 'down' state universally (**Figs. S10(i-j)**).

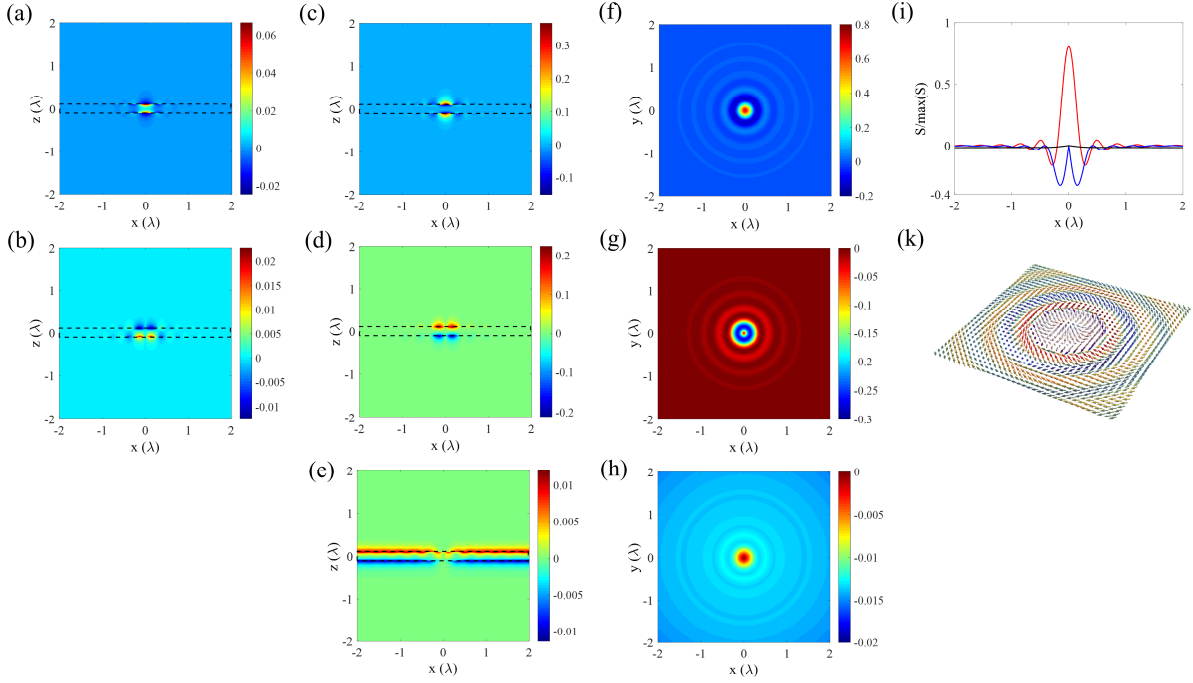

**Fig. S11.** SOC and spin texture of +1-order Bessel-type surface mode at visible frequency. In the metal layer, the (a) z- and (b) radial components of real part of SOC term Re$\mathbf{H}_{SO}$ in xz-plane (y = 0). The (c) z-, (d) radial and (e) azimuthal components of imaginary part of SOC term Im$\mathbf{H}_{SO}$ in xz-plane (y = 0). The (f) z-, (g) radial and (h) azimuthal spin components in xy-plane (z = a/2 − 10nm), and the corresponding (i) 1D contour at y = 0 and (j) spin texture (region: 2λ×2λ). One can find that the spin vectors vary from 'up' state to 'down' state in a period, and thus the skyrmion number $N_{SK} = -1$. However, since an azimuthal spin component exists, it is a manifestation of twisted-type skyrmion texture. This twisted-type skyrmion can be also found in the hyperbolic Metamaterials [S4]. Here, $\lambda = 6.38 \times 10^{-7}$m, which is corresponding to the angular frequency $\omega = 2.95 \times 10^{15}$/s. $a$ is the thickness of metal and here we set $a = 0.2\lambda$.

As the frequency increases (for example, in the near-infrared or visible spectrum), the propagating constant is a complex number and Re($\beta$) > $k_0$ owing to the fact that Re($\varepsilon^m/\varepsilon_0$) < Re($1+\varepsilon^m/\varepsilon_0$) < 0, as indicated by region **II** in **Fig. 1(a)**. In the case, the z-component wavenumber $k_z^m = \sqrt{\beta^2 - (k^m)^2}$ has a same order of magnitudes with the propagating constant $\beta$. Thus, the real and imaginary parts of the z-component SOC term are nonzero ((Re$\mathbf{H}_{SO})_z$ in **Fig. S11(a)** and (Im$\mathbf{H}_{SO})_z$ in **Fig. S11(c)**) and lead to the z-component SAM (**Fig. S11(f)**). In addition, the SOC in the z-direction (**Fig. S11(d)**) results in the radial SAM by $(\hat{\mathbf{n}} \times P_\varphi^A) \cdot \mathbf{S}$ (**Fig. S11(g)**). On the other hand, owing to the complex property of propagation constant or the lossy property of these EM modes, the kinetic Abraham-Poynting momentum in the z-direction appears, and thus the azimuthal SAM component appears since $S_\varphi \propto (\nabla_r \times P_z^A)$ (**Fig. S11(e)**) and thus the spin textures can be considered as the twisted-type skyrmions (skyrmion number $N_{SK} = \pm 1$) (**Fig. S11(i-k)**). Noteworthily, owing to the intrinsic spin-momentum locking of EM field in dispersive medium, the dispersion will cause that the spin-momentum locking is transformed from right hand rule to left hand rule [28], and thus the signs of SAMs in **Fig. S11(c)**/**Fig. S11(d)** and **Fig. S11(f)**/**Fig. S11(g)** are opposite.

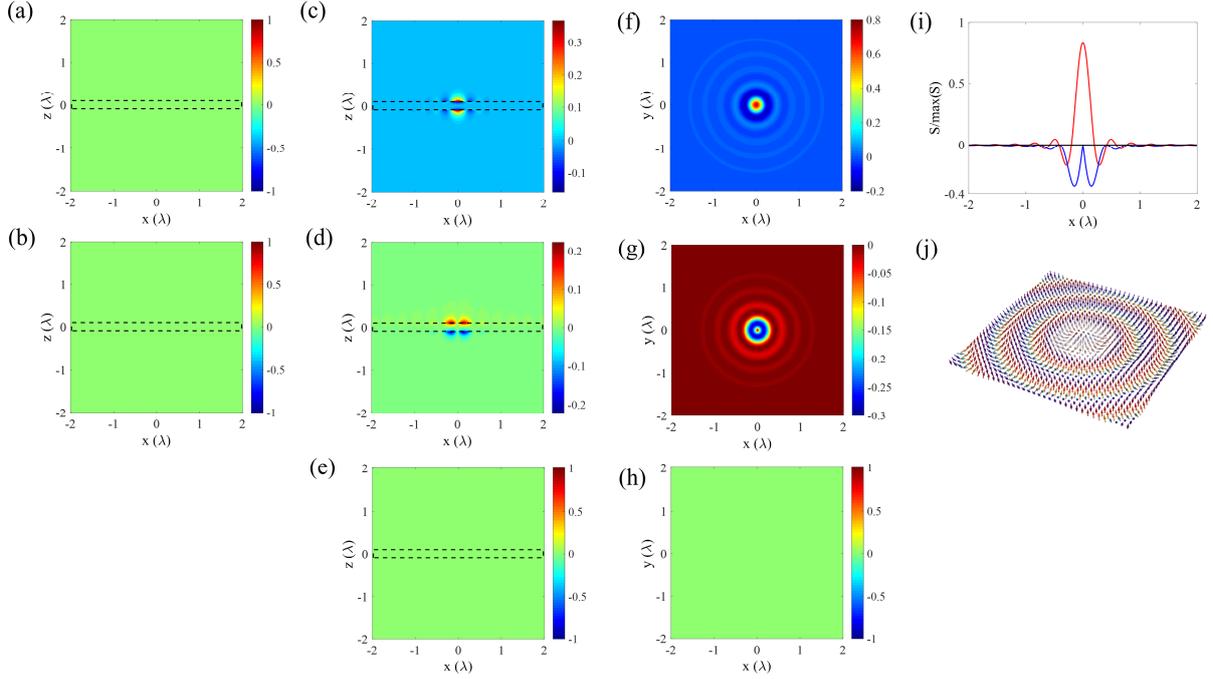

**Fig. S12.** SOC and spin texture of +1-order Bessel-type surface mode at lossless limit. In the metal layer, the (a) *z*- and (b) radial components of real part of SOC term Re**H**$_{SO}$ in *xz*-plane (*y* = 0). The (c) *z*-, (d) radial and (e) azimuthal components of imaginary part of SOC term Im**H**$_{SO}$ in *xz*-plane (*y* = 0). The (f) *z*-, (g) radial and (h) azimuthal spin components in *xy*-plane (z = *a*/2 −10nm), and the corresponding (i) 1D contour at *y* = 0 and (j) spin texture (region: 2*λ*×2*λ*). One can find that, in a period, the spin vectors vary from 'up' state to 'down' state, and thus the skyrmion number $N_{SK} = -1$. The azimuthal spin component is zero in the case, which is a manifestation of Néel-type skyrmion texture. Here, $\gamma = 0$, $\lambda = 6.38 \times 10^{-7}$m, which is corresponding to the angular frequency $\omega = 2.95 \times 10^{15}$/s. *a* is the thickness of metal and here we set $a = 0.2\lambda$.

At the lossless limit, as shown in region **III** of **Fig. 1(a)**, the helical density is zero and the gradient of helical density is also zero (Re**H**$_{SO}$ = 0) (**Fig. S12(a-b)**). However, the *z* component of the imaginary part of SOC term (Im**H**$_{SO}$)$_z$ is nonzero (**Fig. S12(c)**). Thus, there is still the *z*-component SAM. On the other hand, there is only azimuthal kinetic Abraham-Poynting momentum in the layer and thus the azimuthal component of imaginary SOC term disappears (**Fig. S12(d-e)**). Therefore, the configuration of spin topological defect is translated from the twisted type to Néel-type (**Figs. S11(f-j)**). From the intrinsic spin-momentum locking property of EM field in dispersive medium [28], the spin-momentum locking is transformed from right hand rule to left hand rule, and thus the signs of SAMs in **Fig. S12(c)**/**Fig. S12(d)** and **Fig. S12(f)**/**Fig. S12(g)** are opposite.

In the part **IV** of **Fig. 1(a)**, the propagating constant *β* is a pure imaginary number and the propagating mode is forbidden. Thus, no spin topological defect exists in the situation.

As the frequency increases larger than the plasma frequency $\omega > \omega_p$, the metal is transformed to dielectric with relative permittivity smaller than 1. The slot-waveguide-like mode will be present in the layer and the propagation constant is localized beyond the light cone comparing to that of SPP modes (within the light cone), as shown in region **V** of **Fig. 1(a)**. The propagation constant is purely real and equal to the wave number in the medium layer approximatively. Therefore, the decaying factor in the *z*-direction is extremely small and one can consider that the helical density is homogeneous in the layer along the *z*-direction, which causes that the *z*-component of real part of SOC term is zero in the layer (Re**H**$_{SO}$)$_z$ = 0. Thus, the *z*-component SAM is determined by the imaginary part of the SOC term (Im$\mathbf{H}_{SO}$)$_z$. In the case, the kinetic Abraham-Poynting momentum in the $z$-component is nonzero, and the azimuthal SAM appears due to $S_\varphi \propto (\nabla_r \times P_z^A)$ (**Fig. S13(e)**). On the other hand, the radial SAM is given by $(\hat{\mathbf{n}} \times P_\varphi^A) \cdot \mathbf{S}$. Because the outer normal direction of upper interface is inverse to that of the lower interface whereas the azimuthal kinetic Abraham-Poynting momenta are parallel in the upper and lower interfaces, the radial SAM is cancelled out in the layer (**Fig. S13(g)**). This is a manifestation of Bloch-type spin topological defect (**Fig. S13(i-j)**). Moreover, the Bloch-type configuration is stable through the layer, which is one of key achievement of our manuscript. Noteworthily, since the relative permittivity of layer is larger than 0, the intrinsic spin-momentum locking satisfies the right hand rule in the layer, and the sign of SOC term in **Fig. S13(c)/Fig. S13(e)** is consistent with that of SAM in **Fig. S13(f)/Fig. S13(h)**.

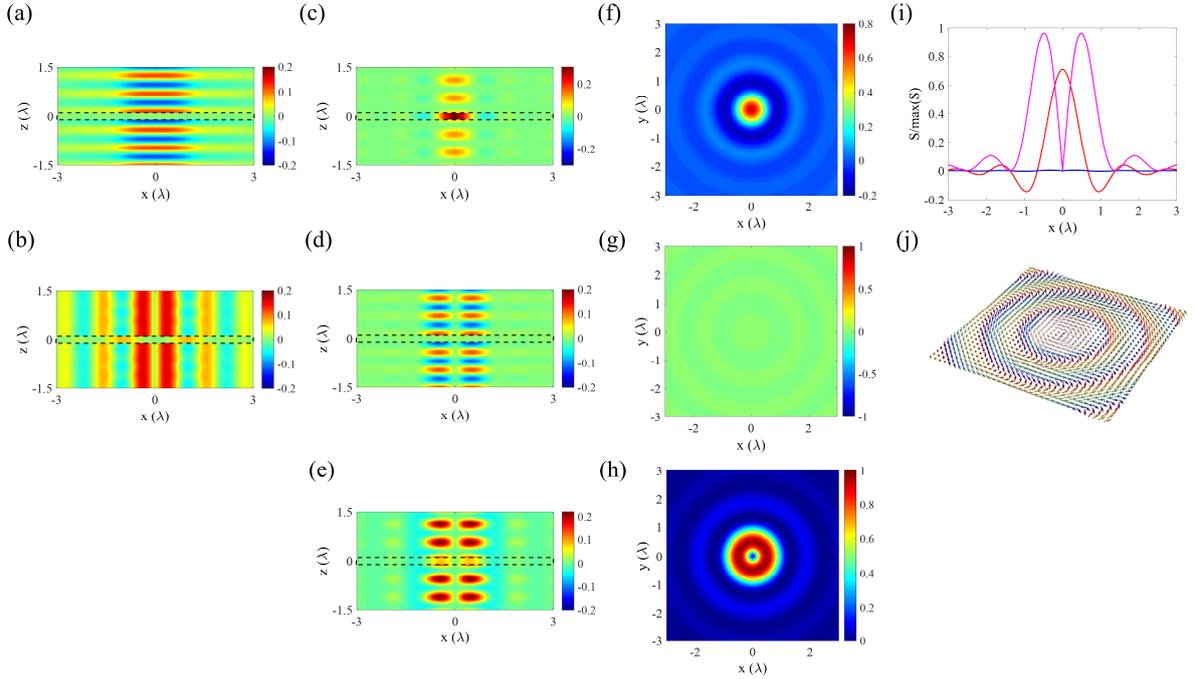

**Fig. S13.** SOC and spin defect of +1-order Bessel-type surface mode at frequency larger than the plasma frequency. In the layer, the (a) $z$- and (b) radial components of real part of SOC term Re$\mathbf{H}_{SO}$ in $xz$-plane ($y = 0$). The (c) $z$-, (d) radial and (e) azimuthal components of imaginary part of SOC term Im$\mathbf{H}_{SO}$ in $xz$-plane ($y = 0$). The (f) $z$-, (g) radial and (h) azimuthal spin components in $xy$-plane ($z = 0$), and the corresponding (i) 1D contour at $y = 0$ and (j) spin texture (region: 8$\lambda$×8$\lambda$). In the case, the slot-waveguide-like modes are present, and the symmetric TE and TM modes with their propagating constants identical are excited. The radial SAM component is cancelled out due to the intrinsic spin-momentum locking property of EM transverse spins in the upper and lower interfaces, which is a manifestation of Bloch-type topological texture. In a period, the spin vectors vary from 'up' state to 'down' state, and thus the skyrmion number $N_{SK} = -1$. Here, $\lambda = 3.33\times10^{-7}m$, which is corresponding to the angular frequency $\omega = 5.66\times10^{15}/s$. $a$ is the thickness of metal and here we set $a = 0.2\lambda$.

For the waveguide mode in **Fig 2(a)**, the $z$-component and radial component of gradient helical density (**Fig. S14(a-b)**) and the three components of vorticities of kinetic Abraham-Poynting momentum (**Fig. S14(c-e)**) appear simultaneously (the $z$-component kinetic Abraham-Poynting momentum is nonzero). Thus, the radial and azimuthal SAM components are existing simultaneously (**Fig. S14(f-h)**) in the layer for the +1-order Bessel-type mode, which is a manifestation of twisted-type spin topological textures (**Fig. S14(i-j)**).

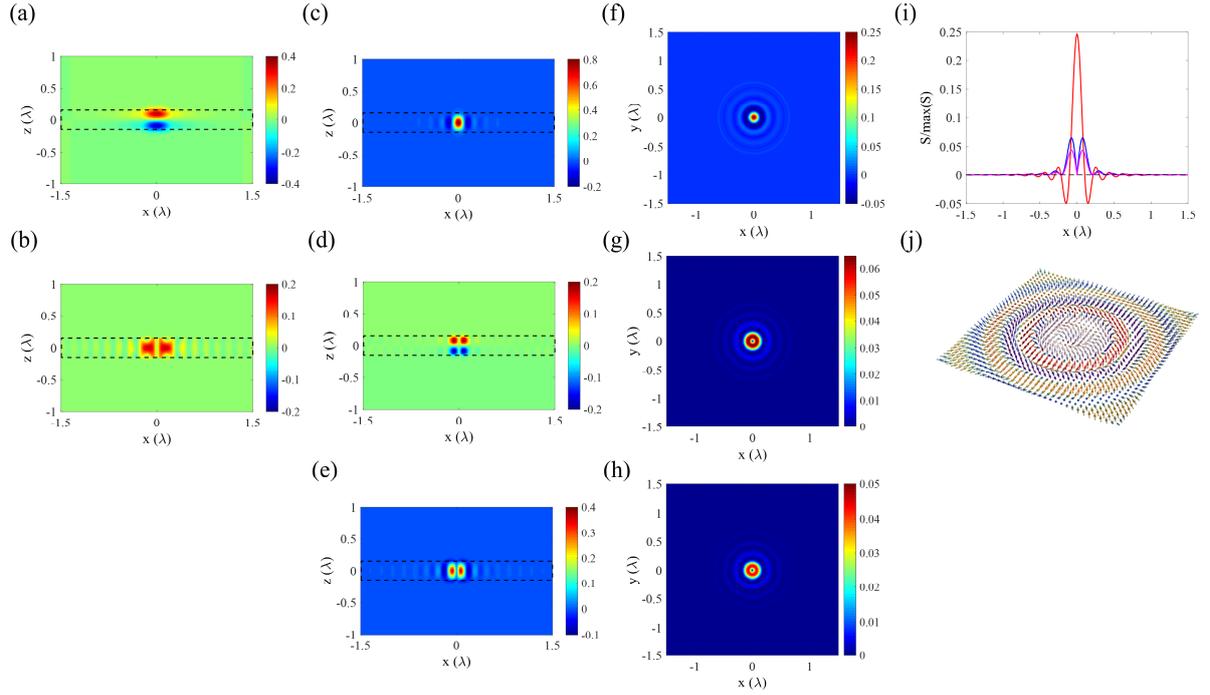

**Fig. S14.** SOC and spin defect of +1-order Bessel-type surface mode at visible frequency. In the layer, the (a) $z$- and (b) radial components of real part of SOC term Re$\mathbf{H}_{SO}$ in $xz$-plane ($y = 0$). The (c) $z$-, (d) radial and (e) azimuthal components of imaginary part of SOC term Im$\mathbf{H}_{SO}$ in $xz$-plane ($y = 0$). The (f) $z$-, (g) radial and (h) azimuthal spin components in $xy$-plane ($z = a/2 - 10$nm), and the corresponding (i) 1D contour at $y = 0$ and (j) spin texture (region: $\lambda \times \lambda$). In a period, the twisted-type spin textures vary from 'up' state to 'down' state. Thus, the skyrmion number is $-1$ ($N_{SK} = -1$). Here, $\lambda = 9.90 \times 10^{-7}$m, which is corresponding to the angular frequency $\omega = 1.90 \times 10^{15}$/s. $a$ is the thickness of metal and here we set $a = 0.3\lambda$.

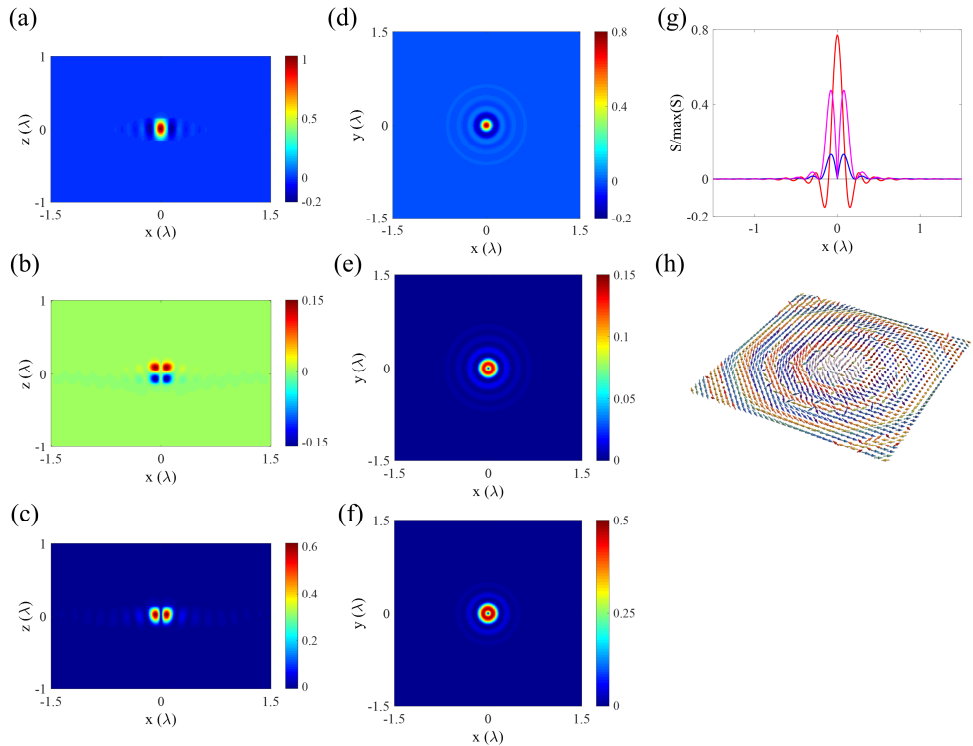

**Fig. S15.** Spin defect of hybrid modes containing the symmetric and antisymmetric +1-order Bessel-type guided wave modes at visible frequency. The (a) $z$-, (b) radial and (c) azimuthal SAM components in $xz$-plane ($y = 0$). In the layer, the (d) $z$-, (e) radial and (f) azimuthal

spin components in *xy*-plane (*z* = *a*/2 −10nm), and the corresponding (g) 1D contour at *y* = 0 and (h) spin texture (region: 1.2$\lambda$×1.2$\lambda$). It can be observed that the skyrmion numbers are unequal to ±1 ($N_{SK} \neq \pm 1$) owing to the mismatch of propagating constants between the symmetric and antisymmetric guided modes, which indicates a topological phase transition. Here, $\lambda = 9.90 \times 10^{-7} m$, which is corresponding to the angular frequency $\omega = 1.90 \times 10^{15}/s$. *a* is the thickness of metal and here we set *a* = 0.3$\lambda$.

In addition, it can be noticed that there will be the symmetric and antisymmetric modes existing for the EM modes within the light cone. When the modes are beyond the light cone ($v_{ph} > c$), only the symmetric modes exist in the structure. More importantly, the propagating constants of symmetric electric and magnetic modes are identical, and hence the topological geometry of the photonic topological textures constructed by these symmetric electric and magnetic EM modes are maintained from the modes' coupling. Whereas if the modes are localized within the light cone ($v_{ph} < c$), the propagating constants of symmetric and antisymmetric modes are different. In this instance, one cannot find a connected boundary to define the topological geometry of this spin texture constructed by the symmetric and antisymmetric EM modes simultaneously (**Fig. S15**). Therefore, we can conclude that the coupling between these symmetric and antisymmetric modes will result in a photonic analogy of topological phase transition for the spin texture.

## VI. EM topological defects in C4 and C6 rotating symmetry

Here, we first give the details of derivations of field distributions for the photonic meron and skyrmion lattices in C4 and C6 rotating symmetries, respectively.

The normal electric field component $E_z = A/\varepsilon \beta^2 \xi$ should fulfill the Helmholtz equation

$$\nabla_\perp^2 \xi(x,y,z) + \beta^2 \xi(x,y,z) = 0, \tag{S79}$$

where the trial solution can be expressed as

$$\xi(x,y,z) = X(x)Y(y)e^{\pm k_z z}. \tag{S80}$$

Here, the +$k_z z$ indicates the field decay to the negative infinity (below the interface) while −$k_z z$ indicates the field decay to the positive infinity (above the interface). By substituting equation (S80) into equation (S79), it can be obtained that

$$\frac{1}{X}\frac{\partial^2 X}{\partial x^2} + \frac{1}{Y}\frac{\partial^2 Y}{\partial y^2} + \beta^2 = 0, \tag{S81}$$

which can be separated into

$$\begin{cases} \frac{1}{X}\frac{\partial^2 X}{\partial x^2} + u\beta^2 = 0 \\ \frac{1}{Y}\frac{\partial^2 Y}{\partial y^2} + v\beta^2 = 0 \end{cases} \tag{S82}$$

with *u* + *v* = 1. The nontrivial solution of expression (S79) is

$$\xi(x,y,z) = \begin{Bmatrix} A\sin(\sqrt{u}\beta x)\sin(\sqrt{v}\beta y) + B\cos(\sqrt{u}\beta x)\cos(\sqrt{v}\beta y) \\ +C\sin(\sqrt{u}\beta x)\cos(\sqrt{v}\beta y) + D\cos(\sqrt{u}\beta x)\sin(\sqrt{v}\beta y) \end{Bmatrix} e^{\pm k_z z}. \tag{S83}$$

If we set

$$L_x = \frac{2n\pi}{\sqrt{u}\beta} = 2\lambda_{sp} \quad n = \text{integer}$$
$$L_y = \frac{2m\pi}{\sqrt{v}\beta} = 2\lambda_{sp} \quad m = \text{integer}$$
(S84)

with $\lambda_{sp} = 2\pi/\beta$, one can get

$$\mu + \nu = \left(\frac{2n\pi}{L_x \beta}\right)^2 + \left(\frac{2m\pi}{L_y \beta}\right)^2 = \lambda_{sp}^2 \left[\left(\frac{n}{L_x}\right)^2 + \left(\frac{m}{L_y}\right)^2\right] = \frac{1}{4}\left[n^2 + m^2\right] = 1,$$
(S85)

and then one can obtain that two groups of solutions

$$n = 0 \quad m = \pm 2 \quad \text{and} \quad n = \pm 2 \quad m = 0.$$
(S86)

Therefore, the nontrivial solution in expression (S83) is converted into

$$\xi(x,y,z) = \{A\cos(\beta x) + B\cos(\beta y) + C\sin(\pm\beta x) + D\sin(\pm\beta y)\}e^{\pm k_z z}$$
$$= \{A'\cos(\beta x) + B'\cos(\beta y) + C'\sin(\beta x) + D'\sin(\beta y)\}e^{\pm k_z z}.$$
(S87)

The parameters in expression (S87) can be calculated further with rotating symmetry with rotating matrix $R_z(\varphi)$.

The rotating symmetry operator can be expressed as

$$R_z(\varphi)\{E_z[R_z(-\varphi)\vec{r}]\hat{z}\} = e^{il\varphi}\{E_z(\vec{r})\hat{z}\}.$$
(S88)

Note here that there is always $R_z(\varphi)\{E_z\hat{z}\} = E_z\hat{z}$ for the normal electric field component.

For the C4 rotational symmetry, the calculated z-component electric field components are given in **Table. S4**.

Table. S4. Calculated z-component electric field component for the meron lattices

| $l = 0$ | $E_z = \frac{A_0}{\varepsilon}\xi = \frac{A_0}{\varepsilon}\{\cos(\beta x) + \cos(\beta y)\}e^{\pm k_z z}$ |
|---|---|
| $l = 1$ | $E_z = \frac{A_1}{\varepsilon}\xi = \frac{A_1}{\varepsilon}\{\sin(\beta y) + i\sin(\beta x)\}e^{\pm k_z z}$ |
| $l = 2$ | $E_z = \frac{A_2}{\varepsilon}\xi = \frac{A_2}{\varepsilon}\{\cos(\beta y) - \cos(\beta x)\}e^{\pm k_z z}$ |
| $l = 3$ | $E_z = \frac{A_3}{\varepsilon}\xi = \frac{A_3}{\varepsilon}\{\sin(\beta y) - i\sin(\beta x)\}e^{\pm k_z z}$ |

As the $l=0$ or $l=2$, the spin-orbit coupling is absence in the cases, the electric field distributions can be regarded as the field meron lattices [26]. While for $l=1$ and $l=3$, the spin vectors can be regarded as the spin meron lattices [26]. Here, we only consider $l = 1$, where the skyrmion number of spin meron lattices is $\pm 1/2$.

For the C6 rotational symmetry, the calculated z-component electric field components are given in **Table. S5**. As the $l=0$ and $l=3$, the spin-orbit coupling is absence, the electric field distributions can be regarded as the field skyrmion lattices [18,20]. While for $l=1$, $l=2$, $l=4$ and $l=5$, the spin vectors can be regarded as the spin skyrmion lattices [26]. Here, we only consider $l=0$, where the skyrmion number of field skyrmion lattices is $\pm 1$.

**Table. S5.** Calculated *z*-component electric field component for the skyrmion lattices

| $l=0$ | $E_z = \dfrac{B_0}{\varepsilon}\xi = \dfrac{B_0}{\varepsilon}\left\{\cos(\beta x) + \cos\left(\dfrac{1}{2}\beta x + \dfrac{\sqrt{3}}{2}\beta y\right) + \cos\left(\dfrac{1}{2}\beta x - \dfrac{\sqrt{3}}{2}\beta y\right)\right\}e^{\pm k_z z}$ |
|---|---|
| $l=1$ | $E_z = \dfrac{B_1}{\varepsilon}\xi = \dfrac{B_1}{\varepsilon}\left\{\dfrac{2}{\sqrt{3}}\sin\left(\dfrac{1}{2}\beta x\right)\cos\left(\dfrac{\sqrt{3}}{2}\beta y\right) + 2i\cos\left(\dfrac{1}{2}\beta x\right)\sin\left(\dfrac{\sqrt{3}}{2}\beta y\right) + \dfrac{2}{\sqrt{3}}\sin(\beta x)\right\}e^{\pm k_z z}$ |
| $l=2$ | $E_z = \dfrac{B_2}{\varepsilon}\xi = \dfrac{B_2}{\varepsilon}\left\{-2i\sin\left(\dfrac{1}{2}\beta x\right)\sin\left(\dfrac{\sqrt{3}}{2}\beta y\right) - \dfrac{2}{\sqrt{3}}\cos\left(\dfrac{1}{2}\beta x\right)\cos\left(\dfrac{\sqrt{3}}{2}\beta y\right) + \dfrac{2}{\sqrt{3}}\cos(\beta x)\right\}e^{\pm k_z z}$ |
| $l=3$ | $E_z = \dfrac{B_3}{\varepsilon}\xi = \dfrac{B_3}{\varepsilon}\left\{\sin(\beta x) - \sin\left(\dfrac{1}{2}\beta x + \dfrac{\sqrt{3}}{2}\beta y\right) - \sin\left(\dfrac{1}{2}\beta x - \dfrac{\sqrt{3}}{2}\beta y\right)\right\}e^{\pm k_z z}$ |
| $l=4$ | $E_z = \dfrac{B_4}{\varepsilon}\xi = \dfrac{B_4}{\varepsilon}\left\{2i\sin\left(\dfrac{1}{2}\beta x\right)\sin\left(\dfrac{\sqrt{3}}{2}\beta y\right) - \dfrac{2}{\sqrt{3}}\cos\left(\dfrac{1}{2}\beta x\right)\cos\left(\dfrac{\sqrt{3}}{2}\beta y\right) + \dfrac{2}{\sqrt{3}}\cos(\beta x)\right\}e^{\pm k_z z}$ |
| $l=5$ | $E_z = \dfrac{B_5}{\varepsilon}\xi = \dfrac{B_5}{\varepsilon}\left\{\dfrac{2}{\sqrt{3}}\sin\left(\dfrac{1}{2}\beta x\right)\cos\left(\dfrac{\sqrt{3}}{2}\beta y\right) - 2i\cos\left(\dfrac{1}{2}\beta x\right)\sin\left(\dfrac{\sqrt{3}}{2}\beta y\right) + \dfrac{2}{\sqrt{3}}\sin(\beta x)\right\}e^{\pm k_z z}$ |

**REFERENCES:**

[S1] L. D. Landau, E. M. Lifshitz, and L. P. Pitaevskii, *Electrodynamics of Continuous Media* (Pergamon, Oxford, 1984).

[S2] P. Shi, H. Li, L. Du, and X. Yuan, "Spin-momentum properties in the paraxial optical system," *ACS Photonics* (2022). DOI: 10.1021/acsphotonics.2c01535

[S3] Anatoly V. Zayats a, Igor I. Smolyaninov b, Alexei A. Maradudin, "Nano-optics of surface plasmon polaritons," *Phys. Rep.* **408**(3–4), 131-314 (2005).

[S4] S. Gan, P. Shi, A. Yang, M. Lin, L. Du, and X. Yuan, "Deep-Subwavelength Optical Spin Textures in Volume Plasmon Polaritons with Hyperbolic Metamaterials," *Adv. Optical Mater.* **11**, 2201986 (2023).